\documentclass[12pt]{iopart}

\usepackage{bm}
\usepackage{hyperref}
\usepackage{iopams}
\usepackage{setstack}

\newcommand{\eg}{\em{eg.}}

\newcommand{\mgut}{M_{\mbox{\scriptsize GUT}}}
\newcommand{\planck}{M_{\mbox{\scriptsize P}}}
\newcommand{\gev}{\mbox{GeV}}

\newcommand{\imag}{\mathrm{i}}
\newcommand{\reals}{\bm{\mathrm{R}}}

\newcommand{\cont}[1]{\mathrm{C}^{{#1}}}
\newcommand{\diracd}{\delta}
\renewcommand{\d}{\mathrm{d}}
\renewcommand{\e}[1]{\mathrm{e}^{{#1}}}
\newcommand{\vect}[1]{\bm{\mathrm{{#1}}}}
\newcommand{\dimension}[1]{[\mathrm{{#1}}]}
\newcommand{\even}[1]{{\mbox{\scriptsize ${#1}$ even}}}
\newcommand{\swaplabel}[2]{[\![\bf{{{#1}}} \leftrightharpoons \bf{{{#2}}}]\!]}

\newcommand{\Prob}{\bm{\mathrm{P}}}
\newcommand{\Probsm}{\Prob}
\newcommand{\Expect}{\bm{\mathrm{E}}}

\newcommand{\bra}[1]{\langle{{#1}}|}
\newcommand{\ket}[1]{|{{#1}}\rangle}
\newcommand{\invac}{\mathrm{in}}
\newcommand{\outvac}{\mathrm{out}}
\newcommand{\inket}{\ket{\invac}}
\newcommand{\outbra}{\bra{\outvac}}
\newcommand{\inbra}{\bra{\invac}}
\newcommand{\outket}{\ket{\outvac}}

\newcommand{\R}{\mathcal{R}}
\newcommand{\Rsp}{\mathsf{R}}
\newcommand{\Rsm}{\bar{\Rsp}}
\newcommand{\A}{\mathcal{A}}
\newcommand{\Asq}{\alpha}
\newcommand{\fnl}{f_{\mathrm{NL}}}
\newcommand{\Gauss}{\Gamma}
\newcommand{\correction}{\Upsilon}
\newcommand{\Gaussm}{\Gauss}
\newcommand{\pregauss}{\omega}
\newcommand{\window}{\mathcal{W}}
\newcommand{\kmax}{k_{H}}
\newcommand{\spect}{\varrho}
\newcommand{\fluct}{\epsilon}
\newcommand{\ps}{\mathcal{P}}
\newcommand{\Jtri}{J_{\mathrm{tri}}}

\begin{document}
\title[Non-Gaussian corrections to the curvature perturbation]
{Non-Gaussian corrections to the probability distribution of the
curvature perturbation from inflation}
\date{\today}
\author{David Seery$^1$ and J. Carlos Hidalgo$^{1,2}$}
\vspace{3mm}
\address{$^1$ Astronomy Unit, School of Mathematical Sciences\\
  Queen Mary, University of London\\
  Mile End Road, London E1 4NS\\
  United Kingdom}
\vspace{2mm}
\address{$^2$ Instituto de Astronom\'{\i}a \\
  Universidad Nacional Aut\'{o}noma de M\'{e}xico \\
  AP 70-264, Distrito Federal 04510 \\
  M\'{e}xico}
\eads{\mailto{D.Seery@qmul.ac.uk}, \mailto{C.Hidalgo@qmul.ac.uk}}
\vspace{2mm}
\submitto{JCAP}
\begin{abstract}
We show how to obtain the probability density function for the amplitude of
the curvature perturbation, $\R$, produced during an epoch of slow-roll,
single-field inflation,
working directly from $n$-point correlation functions of $\R$.
These $n$-point functions are the usual output of quantum field theory
calculations, and as a result we
bypass approximate statistical arguments based on the central limit theorem. Our
method can be extended to deal with arbitrary forms of non-Gaussianity,
appearing at any order in the $n$-point hierarchy.
We compute the probability density for the total smoothed perturbation
within a Hubble volume, $\fluct$, and for the spectrum of $\fluct$.
When only the two-point function is retained, exact Gaussian statistics
are recovered. When the three-point function is taken into account, we
compute explicitly the leading slow-roll correction to the Gaussian
result.
\vspace{3mm}
\begin{flushleft} \textbf{Keywords}:
Inflation,
Cosmological perturbation theory,
Physics of the early universe
\end{flushleft}
\end{abstract}
\maketitle

\section{Introduction}
\label{sec:intro}
There is now a good deal of observational evidence that the generic predictions
of the inflationary scenario are realised in the spectrum of density
perturbations in our universe \cite{wmap2006-params,wmap2006-temp,
wmap2006-pol,boomerang}.
For slow-roll inflation driven by a scalar $\phi$, these predictions are:
\begin{enumerate}
  \item a nearly scale-invariant spectrum of fluctuations on all scales
  accessible to cosmological observation; \label{intro:cond:a}
  \item for inflation near the theoretically motivated energy scale
  $\mgut \simeq 10^{16} \, \gev$, these fluctuations should
  have magnitude $\delta \rho/\rho \simeq 10^{-5}$; and
  \label{intro:cond:b}
  \item the fluctuation spectrum should exhibit Gaussian statistics, in the
  sense that the probability distribution of the density fluctuation should be
  approximately normally distributed. \label{intro:cond:c}
\end{enumerate}
(For a review of the inflationary
paradigm and its predictions, see \emph{eg}., Refs. \cite{liddle-lyth,peacock}.)
Predictions (\ref{intro:cond:a}) and (\ref{intro:cond:b}) can be obtained
using the standard techniques of quantum field theory.
This calculation is now classical \cite{liddle-lyth-paper,liddle-lyth,peacock}
and relies only on the fact that the vacuum fluctuation of a scalar field
in de Sitter space with Hubble parameter $H$ is roughly $\delta \phi =
H/2\pi$  \cite{bardeen-steinhardt-turner,guth-pi,hawking-fluct,
starobinsky-fluct}.
The third prediction---that the fluctuation spectrum
is Gaussian---is less transparent.
It follows from the fact that the inflaton perturbation, which is commonly
expressed in terms of the so-called comoving curvature perturbation%
\footnote{$\R(t,\vect{x})$ expresses the relative expansion
of a given local neighbourhood of the universe with respect to the background,
in a gauge where observers in
free-fall with the expansion see no net momentum flux.
This is the so-called comoving gauge. In particular, during
inflation the scalar field fluctuation is zero on comoving slices.
The metric in this gauge is thus
\begin{equation}
  \d s^2 = - N^2(t,\vect{x}) \, \d t^2 + a^2(t) \e{2\R(t,\vect{x})} \gamma_{ij}
  (\d x^i + N^i(t,\vect{x}) \, \d t)( \d x^j + N^j(t,\vect{x}) \, \d t) ,
\end{equation}
where $\gamma_{ij}$ is the metric on unperturbed spatial slices, which we take
to be flat and is subject to the condition $\det \gamma = 1$;
and $a(t)$ is the unperturbed scale factor of the universe.
The functions
$N$ and $N^i$ are the so-called lapse function and shift vector, and are
determined by algebraic constraint equations
in terms of the matter content and the metric fields
($\R$, $a$ and $\gamma$), once the gauge is fixed.
Our sign convention for the metric is ``mostly plus'', $(-,+,+,+)$.
The advantage of working with the perturbation
$\R(t,\vect{x})$ is that it mixes scalar fluctuations from the metric and
matter sectors in a gauge invariant way \cite{mfb}.}
\cite{wands-malik} $\R$, is treated as a free field,
\begin{equation}
  \label{intro:freer}
  \R(t,\vect{x}) = \int \frac{\d^3 k}{(2\pi)^3} \; \R(t,\vect{k}) \e{\imag
  \vect{k}\cdot\vect{x}} ,
\end{equation}
where there is no coupling between the $\R(t,\vect{k})$ for different
$\vect{k}$.
With this understanding, Eq.~\eref{intro:freer}
means that $\R$ does not interact either with itself, or any other particle
species in the universe.
The real-space field
$\R(t,\vect{x})$ is obtained by summing an infinite number of independent,
identically distributed, uncorrelated oscillators.
Under these circumstances the
Gaussianity of $\R(t,\vect{x})$ follows from the central limit theorem
\cite{bardeen-bond},
given reasonable assumptions about the individual distributions of the
$\R(t,\vect{k})$. The exact form of the distributions
of the $\R(t,\vect{k})$ is
mostly irrelevant when making statements about the inflationary density
perturbation.

In conventional quantum field theory, all details
of $\R$ and its interactions are encoded in the $n$-point correlation functions
of $\R$ (or their Fourier transforms), written
$\outbra \R(t_1,\vect{x}_1) \cdots \R(t_n,\vect{x}_n) \inket$.
Working in the Heisenberg picture, where the fields carry time dependence but
the states $\{ \inket$, $\outket \}$ do not, 
these functions express the amplitude for the early-time
vacuum $\inket$ to evolve into the late-time vacuum $\outket$ in the
presence of the fields $\R(t_i,\vect{x}_i)$.
Given the $n$-point functions for all $n$ at arbitrary $\vect{x}$ and $t$,
one can determine $\R(t,\vect{x})$ \cite{streater-wightman}, at least in
scattering theory.
In the context of the inflationary density perturbation, these vacuum evolution
amplitudes are not directly relevant. Instead, one is interested in the
equal time expectation values
$\inbra \R(t,\vect{x}_1) \cdots \R(t,\vect{x}_n) \inket$, which can be
used to measure gravitational particle creation
out of the time-independent early vacuum $\inket$
during inflation. These expectation values are calculated using
the so-called ``closed time path formalism,''
which was introduced by Schwinger \cite{schwinger-ctp} (see also
\cite{calzetta-hu,jordan,dewitt,hajicek}).
In this formalism there is a doubling of degrees of freedom, which is also
manifest in finite temperature calculations \cite{lebellac,rivers}.
This method has recently been used \cite{weinberg-corrl,weinberg-corrl-a,sloth}
to extend the computation of the correlation functions of $\R$ beyond
tree-level.

Knowledge of the expectation values of $\R$ in the state $\inket$ is sufficient
to predict a large number of
cosmological observables, including the power spectrum of the
density perturbation generated during inflation
\cite{guth-pi,hawking-fluct}, and the two- and three-point
functions of the cosmic microwave background (CMB)
\cite{hu-sugiyama,hu-trispectrum,komatsu-spergel,kogo-komatsu,
okamoto-hu,babich-creminelli,
babich-zaldarriaga,babich,liguori-hansen,cabella-hansen,creminelli-nicolis}
temperature anisotropy. Because they are
defined as expectation values in the quantum vacuum these observables all
have the interpretation of ensemble averages, as will be discussed in more
detail below.
On the other hand, it is sometimes required to know the probability that
fluctuations of some given magnitude occur in the perturbation $\R$
\cite{press-schechter,bardeen-bond,peacock-heavens}.
This is not a question about ensemble averages, but instead is concerned
with the probability measure on the ensemble itself.
As a result, such information cannot easily be
obtained from inspection or simple manipulation of the $n$-point functions.

For example, if we know by some a priori means that $\R$ is free,
then the argument given above based on the
central limit theorem implies that at any position $\vect{x}$, the probability
of fluctuations in $\R$ of size $\fluct$ must be
\begin{equation}
  \label{intro:gaussian}
  \Prob(\mbox{$\R$ has fluctuations of size $\fluct$})
  \simeq \frac{1}{\sqrt{2\pi} \sigma}
  \exp \left( - \frac{\fluct^2}{2\sigma^2} \right) ,
\end{equation}
where the variance in $\R$ is
\begin{equation}
  \label{intro:variance}
  \sigma^2 = \langle \R(t,\vect{x})^2 \rangle =
  \int \d \ln k \; \ps(k) .
\end{equation}
The quantity $\ps(k)$ is the so-called dimensionless power spectrum,
which is defined in terms of the two-point function of $\R$,
calculated from the quantum field theory in-vacuum:
\begin{equation}
  \label{intro:twopt}
  \inbra \R(t,\vect{k}_1) \R(t,\vect{k}_2) \inket
  = (2\pi)^3\diracd(\vect{k}_1 + \vect{k}_2)\frac{2\pi^2}{k_1^3} \ps(k_1) .
\end{equation}
This is the only relevant observable, because
it is a standard property of free fields that
all other non-vanishing correlation functions can be
expressed in terms of the two-point function \eref{intro:twopt}, and hence
the power spectrum.
In practice, in order to give a precise meaning to \eref{intro:gaussian},
it would be necessary
to specify what it means for $\R$ to develop fluctuations of size $\fluct$,
and whether it is the fluctuations in the microphysical field
$\R$ or some smoothed field $\bar{\R}$ which are measured.
These details affect the exact expression \eref{intro:variance} for the
variance of $\fluct$.

The average in Eq. \eref{intro:twopt},
denoted by $\inbra \cdots \inket$, is the expectation value in the
quantum in-vacuum. To relate this abstract expectation value
to real-world measurement probabilities, one introduces a notional ensemble of
possible universes, of which the present universe and the density fluctuation
that we observe is only one possible realization (see, \emph{eg}.,
\cite{lyth-curvaton-ng}).
However,
for ergodic processes, we may freely trade ensemble averages for volume
averages. If we make the common supposition that the inflationary density
perturbation is indeed ergodic, then we expect the volume average of the density
fluctuation to behave like the ensemble average: the universe may contain
regions where the fluctuation is atypical, but with high probability
most regions contain fluctuations of root mean square (rms)
amplitude close to $\sigma$.
Therefore the probability distribution on the ensemble, which is encoded
in Eq. \eref{intro:twopt}, translates to a probability distribution on
smoothed regions of order the horizon size within our own universe.

In order to apply the above analysis, it was necessary to know in advance that
$\R$ was a free field. This knowledge allowed us to use the
central limit theorem to connect the correlation functions of
$\R$ with the probability distribution \eref{intro:gaussian}.
The situation in the real universe is not so simple. In particular,
the assumption that during inflation $\R$ behaves as a free field, and
therefore that the oscillators $\R(\vect{k})$ are uncorrelated and
independently distributed, is only approximately correct.
In fact, $\R$ is subject to
self-interactions and interactions with the other constituents of the universe,
which mix $\vect{k}$-modes. Consequently, the oscillators
$\R(\vect{k})$ acquire some phase correlation and are no longer independently
distributed.
In this situation the central limit theorem gives
only approximate information concerning the
probability distribution of $\R(\vect{x})$, and it is necessary to
use a different method to connect the correlation functions of $\R$ with
its probability distribution.

In this paper we give a new derivation of the probability distribution
of the amplitude of fluctuations in $\R$ which directly connects
$\Prob(\fluct)$ and the correlation
functions $\langle \R(\vect{k}_1) \cdots \R(\vect{k}_n) \rangle$, without
intermediate steps which invoke the central limit theorem or other
statistical results.
When the inflaton is treated as a free field, our method
reproduces the familiar prediction \eref{intro:gaussian} of Gaussian statistics.
When the inflaton is \emph{not} treated as a free field,
the very significant advantage of our
technique is that it is possible to directly calculate the corrections to
$\Prob(\fluct)$. Specifically,
the interactions of $\R$ can be measured by the departure of the correlation
functions from the form they would take if $\R$ were free.
Therefore, the first corrections to the free-field approximation
are contained in the three-point function, which is exactly zero when there
are no interactions.

The three-point function for single-field, slow-roll inflation
has been calculated by Maldacena \cite{maldacena-nongaussian},
whose result can be expressed in the form \cite{seery-lidsey-a}
\begin{equation}
  \label{intro:threept}
  \langle \R(\vect{k}_1) \R(\vect{k}_2) \R(\vect{k}_3) \rangle =
  4 \pi^4 (2\pi)^3 \diracd(\sum_i \vect{k}_i) \frac{\bar{\ps}^2}
  {\prod_i k_i^3} \A(k_1,k_2,k_3) ,
\end{equation}
where $\A$ is one-half%
\footnote{In Maldacena's normalization, the numerical prefactor in
Eq.~\eref{intro:threept} is not consistent with the square of
the two-point function,
Eq.~\eref{intro:twopt}. We choose $\A$ so that the prefactor becomes
$4\pi^2(2\pi)^3$. This normalization of Eq.~\eref{intro:threept}
was also employed in Refs. \cite{seery-lidsey,seery-lidsey-a},
although the distinction from Maldacena's $\A$ was not pointed out explicitly.
Throughout we work in units where the reduced Planck mass,
defined by $\planck^2 = (8\pi G)^{-1}$, is set to unity. If necessary,
a finite Planck mass can be restored in formulas such as
\eref{intro:threept} by dimensional analysis.}
Maldacena's $\A$-function \cite{maldacena-nongaussian} and
$\bar{\ps}^2$ (to be defined later) measures
the amplitude of the spectrum when the $\vect{k}_i$ crossed the horizon.
(For earlier work, see Refs. \cite{falk-rangarajan,gangui-lucchin,
pyne-carroll,acquaviva-bartolo}. The present situation is reviewed in Ref.
\cite{bartolo-matarrese-review}.)
This result has since been extended to cover the non-Gaussianity produced
during slow-roll inflation with an arbitrary number of fields
\cite{maldacena-nongaussian,seery-lidsey-a,
creminelli,lyth-rodriguez,lyth-rodriguez-a,lyth-zaballa,
zaballa-rodriguez,vernizzi-wands},
preheating \cite{enqvist-jokinen,enqvist-jokinen-a,jokinen-mazumdar},
models where the dominant non-Gaussianity is produced by a light scalar which
is a spectator during inflation \cite{boubekeur-lyth,alabidi-lyth,
lyth-curvaton-ng},
and alternative models, often involving a small speed of sound for the
inflaton perturbation
\cite{seery-lidsey,
alishahiha-silverstein,calcagni-nongaussian,arkani-hamed-creminelli,
creminelli}.
For single-field, slow-roll inflation,
the self-interactions of $\R$ are suppressed by
powers of the slow-roll parameters. This means that the correction to Gaussian
statistics is not large. In terms of the $\A$-parametrized three-point function
\eref{intro:threept}, this is most commonly expressed by writing
\begin{equation}
  \fnl = - \frac{5}{6} \frac{\A}{\sum_i k_i^3} = \Or(\mbox{slow roll}) ,
\end{equation}
where $\fnl$ \cite{komatsu-spergel,verde-wang} expresses the relative
contribution
of a non-Gaussian piece in $\R$, viz, $\R = \R_g - \frac{3}{5} \fnl \star
\R_g^2$, and $\R_g$ is a Gaussian random field. (There are
differing sign conventions for $\fnl$ \cite{lyth-curvaton-ng}.)
In models with more degrees of freedom it is expected to be possible to
obtain very much larger non-Gaussianities, perhaps with $\fnl \sim 10$
\cite{rigopoulos-shellard,
rigopoulos-shellard-vantent,rigopoulos-shellard-vantent-b,
rigopoulos-shellard-vantent-c,boubekeur-lyth,lyth-rodriguez-a,
vernizzi-wands},
although no unambiguous concrete example yet exists except where the
non-Gaussianity is generated during preheating
\cite{enqvist-jokinen,jokinen-mazumdar}. If the inflationary
perturbation has a speed of sound different from unity then large
non-Gaussianities may also appear \cite{seery-lidsey}, although in this case
it is difficult to simultaneously achieve scale invariance.
The current observational constraint can approximately be expressed as
$|\fnl| \lesssim 100$ \cite{wmap2006-params}. In the absence of a detection,
the forthcoming Planck Explorer mission may tighten this constraint to
$|\fnl| \lesssim 3$ \cite{komatsu-spergel,liguori-hansen}.

Non-Gaussian probability distributions have been studied
previously by several authors.
The closest to the method developed in this paper include
that of Matarrese, Verde \& Riotto
\cite{matarrese-verde}, who worked with a path integral expression
for the density fluctuation smoothed on a scale $R$
(which they denoted `$\delta_R$'); and that of Bernardeau \& Uzan
\cite{bernardeau-uzan,bernardeau-uzan-a}.
The latter analysis has some features in common with our own,
being based on the cumulant generating function,
and moreover since the expression for the probability density in
Refs. \cite{bernardeau-uzan,bernardeau-uzan-a} is expressed as a Laplace
transform. Our final expression, Eq.~\eref{hubble:prob}, can be interpreted
as a Fourier integral, viz \eref{genfunc:prob}, which (loosely speaking) can be
related to a Laplace integral via a Wick transformation.
Despite these similarities,
the correspondence between the two is complicated because
Refs. \cite{bernardeau-uzan,bernardeau-uzan-a} work in a multiple-field
picture and calculate a probability density only for the isocurvature
field `$\delta s$', which acquires its non-Gaussianity via a mixing of
isocurvature and adiabatic modes long after horizon exit. This contrasts with
the situation in the present paper, where we restrict ourselves to a
single-field scenario and compute the probability density for the
adiabatic mode $\R$, which
would be orthogonal to $\delta s$ in field space and whose non-Gaussianity
is generated exactly at horizon exit.

In the older literature it is more common to deal with the density fluctuation
$\delta\rho/\rho$ measured on comoving slices, as an alternative to the
curvature perturbation $\R$.
For slowly varying fields, on large scales,
$\R$ and $\delta\rho$ can be related via the rule
(Eq. (25) of Ref. \cite{lyth-classicality}),
\begin{equation}
  \left( \frac{aH}{k} \right)^2 \frac{\delta\rho}{\rho} =
  - \left( \frac{3}{2} + \frac{1}{1+\omega} \right)^{-1} \R ,
\end{equation}
which is valid outside the horizon, and to first order in cosmological
perturbation theory, for a fluid with equation of state $p = \omega\rho$.
(One may use the $\delta N$ formalism to go beyond
leading order, but to obtain results valid on sub-horizon scales one must
use the full Einstein equations directly; see, {\eg},
\cite{langlois-vernizzi-a,langlois-vernizzi-b}.)
For fluctuations on the Hubble scale, where
$k \simeq aH$, this means $|\R| \simeq \delta\rho/\rho$, so $\R$ provides
a useful measure of the density fluctuation on such scales.
In virtue of this relationship with the density fluctuation,
the probability distribution $\Prob(\fluct)$ is an important
theoretical tool, especially in studies of structure formation. For example,
it is the principal object in the Press--Schechter formalism
\cite{press-schechter}. As a result, there are important
reasons why knowledge of the detailed form of the probability distribution of
$\fluct$ is important, and not merely the approximate answer provided by
the central limit theorem.

Firstly, large collapsed objects,
such as primordial black
holes (PBHs) naturally form on the high-$\fluct$ tail of the distribution
\cite{carr,carr-hawking}.
Such large fluctuations are extremely rare. This means that a small change in
the probability density for $|\fluct| \gg 0$ can make a large difference in the
mass fraction of the universe which collapses into PBHs
\cite{bullock-primack,ivanov}. Thus
one may hope to probe it using well-known and
extremely stringent constraints
on PBH formation in the early universe
\cite{carr-constraints,carr-lidsey-constraints}.
The corrections calculated in this paper are
therefore not merely of theoretical interest, but relate directly to
observations, and have the potential to sharply discriminate between models
of inflation.

Secondly, as described above, although the non-Gaussianities produced
by single-field, slow-roll inflation are small, this is not mandatory.
In models where non-Gaussianities are large,
it will be very important to account for the effect of non-Gaussian fluctuations
on structure formation
\cite{verde-wang,matarrese-verde,verde-jimenez,verde-trispectrum}.
The formalism presented in this paper provides
a systematic way to obtain such predictions, extending the analysis given in
Ref. \cite{matarrese-verde}.

The outline of this paper is as follows.
In Section~\ref{sec:measure} we obtain the probability measure on the ensemble
of possible fluctuations. This step depends on the correlation functions
of $\R$. In Section~\ref{sec:harmoniccurv}, we discuss
the decomposition of $\R$ into harmonics. This is a technical step, which is
necessary in order to write down a path integral for $\Prob(\fluct)$.
First, $\R$ is decomposed in Section~\ref{sec:harmonic}. The path integral
measure is written down in Section~\ref{sec:pathmeasure}, and
in Section~\ref{sec:condition} we give a
precise specification of $\fluct$, which measures the size of fluctuations.
We distinguish two interesting cases,
a ``total fluctuation'' $\fluct$,
which corresponds to $\R$ (or approximately $\delta\rho/\rho$) smoothed over
regions the size of the Hubble volume, and the ``spectrum'' $\spect(k)$,
which describes the contributions to $\fluct$ from regions of the
primordial power spectrum around the scale described by wavenumber $k$.
In Section~\ref{sec:probdelta} we
evaluate $\Prob(\fluct)$. We give the calculation for the Gaussian case first,
in Section~\ref{sec:gaussian}, in order to clearly explain our method with
a minimum of technical distractions. This is
followed in Section~\ref{sec:nongaussian} by the same calculation
but including non-Gaussian corrections which follow from a non-zero
three-point function.
In Section~\ref{sec:probrho} we calculate $\Prob[\spect(k)]$.
Finally, we state our conclusions in Section~\ref{sec:conclude}.

\section{The probability measure on the ensemble of $\R$}
\label{sec:measure}
Our method is to compute the probability measure $\Prob_t[\Rsp]$ on the
ensemble of realizations of the curvature perturbation $\Rsp(\vect{x})$,
which we define to be the value of $\R(t,\vect{x})$ at some fixed time $t$.
This
probability measure is a natural object in the Schr\"{o}dinger approach to
quantum field theory, where the elementary quantity is the wavefunctional
$\Psi_t[\Rsp]$, which is
related to $\Prob_t[\Rsp]$ by the usual rule
of quantum mechanics, that $\Prob_t[\Rsp] \propto |\Psi_t[\Rsp]|^2$. Once
the measure $\Prob_t[\Rsp]$ is known, we can directly calculate
(for example) $\Prob_t(\fluct)$
by integrating over all $\Rsp$ that produce fluctuations of amplitude
$\fluct$. Although the concept of a probability
measure on $\Rsp$ may seem like a rather formal object, the Schr\"{o}dinger
representation%
\footnote{The Schr\"{o}dinger representation is briefly discussed in,
for example, Refs \cite{polchinski,visser}. A brief introduction to
infinite-dimensional probability measures is given in Ref.
\cite{albeverio}.}
of quantum field theory is entirely equivalent to the more familiar formulation
in terms of a Fock space. Indeed, a similar procedure has been discussed by
Ivanov \cite{ivanov}, who calculated the probability measure on a stochastic
metric variable $a_{\mathrm{ls}}(\vect{x})$ which can be related to our
$\R(\vect{x})$. Although the approaches are conceptually
similar, our method is substantially
different in detail. In particular, the present
calculation is exact in the sense that we make no reference to the stochastic
approach to inflation, and therefore are not obliged to introduce a
coarse-graining approximation. Moreover, Ivanov's analysis appeared before
the complete non-Gaussianity arising from $\R$-field interactions around
the time of horizon crossing had been calculated \cite{maldacena-nongaussian},
and therefore did not include this effect.

\subsection{The generating functional of correlation functions}
The expectation values $\langle \R(\vect{x}_1) \cdots \R(\vect{x}_2)
\rangle$ in the vacuum $\inket$
at some fixed time $t$ can be expressed in terms of a
Schwinger--Keldysh path integral,%
\footnote{Henceforth, we use the notation $\langle \cdots \rangle$ to
mean expectation values in the in-vacuum, and no longer write
$\inket$ explicitly where this is unambiguous.}
\begin{eqnarray}
  \fl\nonumber
  \langle \R(t,\vect{x}_1) \cdots \R(t,\vect{x}_n) \rangle = \\
  \label{genfunc:sk}
  \int [\d\R_- \d\R_+]_{\inket}^{\R_{+}(t,\vect{x}) = \R_-(t,\vect{x})}
  \R(t,\vect{x}_1) \cdots \R(t,\vect{x}_n)
  \exp\left( \imag S[\R_+] - \imag S[\R_-] \right) .
\end{eqnarray}
(For details of the Schwinger--Keldysh or ``closed time path''
formalism, see Refs.
\cite{calzetta-hu,jordan,weinberg-corrl,lebellac,hajicek,rivers}.)
In cosmology we are generally interested in $\R$ evaluated at different
spatial positions on the same $t$-slice, so we have set all the $t$ equal
in \eref{genfunc:sk}.
The path integral is taken over all fields $\R$ which begin in a configuration
corresponding to the vacuum $\inket$ at past infinity, and
$S[\R]$ is the action for the fluctuation $\R$, which is given
to third order in $\R$ in Refs.
\cite{maldacena-nongaussian,seery-lidsey,seery-lidsey-a}.

An expression equivalent to Eq.~\eref{genfunc:sk} can be given in terms of the
``equal time'' generating functional $Z_t[\eta]$ at time $t$, defined by
\begin{equation}
  \fl\label{genfunc:def}
  Z_t[\eta] = \int [\d\Rsp] \int [\d\R_- \d\R_+]_{\inket}^
  {\R_{\pm}(t,\vect{x}) = \Rsp(\vect{x})}
  \exp\left( \imag S[\R_+] - \imag S[\R_-] +
  \imag \int_{\Sigma_t} \d^3 x \; \Rsp(\vect{x})\eta(\vect{x}) \right) ,
\end{equation}
where $\eta$ is some arbitrary source field and $\Sigma_t$ is a spatial
slice at coordinate time $t$. This generating functional is
\emph{not} the usual
one, which would generate correlation function at any given set of times,
and not the common time $t$ which appears in
\eref{genfunc:sk}--\eref{genfunc:def}.
The equal-time
correlation functions $\langle \R(t,\vect{x}_1) \cdots \R(t,\vect{x}_n)
\rangle$ are recovered from $Z_t[\eta]$ by functional differentiation,
\begin{equation}
  \label{genfunc:corrls}
  \langle \R(t,\vect{x}_1) \cdots \R(t,\vect{x}_n) \rangle =
  \left.\frac{1}{\imag^n} \frac{\delta}{\delta \eta(\vect{x}_1)} \cdots
  \frac{\delta}{\delta \eta(\vect{x}_n)} \ln Z_t[\eta] \right|_{\eta = 0} .
\end{equation}
Up to normalization,
this is merely the rule for functional Taylor coefficients, so it
is straightforward to invert Eq. \eref{genfunc:corrls} for $Z_t[\eta]$.
We obtain%
\footnote{This construction is somewhat similar
to the dS/CFT calculations outlined in Refs. \cite{maldacena-nongaussian,schaar,
larsen-schaar,seery-lidsey-dscft}.
In these calculations one constructs $|\Psi|^2$
from an expression of the same form as \eref{genfunc:reconstruct}, but
expressed in terms of an operator $\mathcal{O}$ which is the holographic
dual of the bulk field $\R$. The dS/CFT prescription relates the correlators
of $\mathcal{O}$ reciprocally to those of $\R$. In this paper we 
do not make any use of holographic
arguments, but the same reciprocal relationship emerges through the
identification of $Z_t[\eta]$ with the Fourier transform of $|\Psi|^2$.
The normal rules of Fourier transforms show that the transform of a
Gaussian is another Gaussian with a reciprocal coefficient,
$\widetilde{\e{-a x^2/2}}
\overset{\mbox{\scriptsize Fourier}}{\longrightarrow}
\e{-k^2/2a}$.}
\begin{equation}
  \fl\label{genfunc:reconstruct}
  Z_t[\eta] = \exp \sum_{n=0}^\infty \frac{\imag^n}{n!}
  \int \cdots \int \d^3 x_1 \cdots \d^3 x_n \; \eta(\vect{x}_1)
  \cdots \eta(\vect{x}_n) \langle \R(t,\vect{x}_1) \cdots \R(t,\vect{x}_n)
  \rangle ,
\end{equation}

Eq. \eref{genfunc:def} for the generating functional
can be rewritten in a suggestive way. We define
the wavefunctional at time $t$, $\Psi_t[\Rsp]$, as
\begin{equation}
  \Psi_t[\Rsp] = \int [\d\R]_{\inket}^{\R(t,\vect{x}) = \Rsp(\vect{x})}
  \exp\left( \imag S[\R] \right) .
\end{equation}
This definition is simply the functional generalization of the
familiar quantum-mechanical
wavefunction. It expresses the amplitude for the field $\R(t,\vect{x})$ to have
the spatial configuration $\Rsp(\vect{x})$ at time $t$, given the
boundary condition that $\R$ started in the vacuum state in the far past.
In terms of $\Psi_t[\Rsp]$, the generating functional can be rewritten
\begin{equation}
  \fl\label{genfunc:genfunctional}
  Z_t[\eta] = \int [\d \Rsp] \;  \Psi_t[\Rsp]^\dag \Psi_t[\Rsp]
  \exp\left( \imag \int \d^3 x \; \Rsp(\vect{x}) \eta(\vect{x}) \right)
  = \widetilde{|\Psi_t[\Rsp]|^2} \propto \widetilde{\Prob[\Rsp]} ,
\end{equation}
where a tilde denotes a (functional) Fourier transform, and $\dag$
denotes Hermitian conjugation.
Eq.~\eref{genfunc:genfunctional} implies that $Z_t[\eta]$ is the complementary
function \cite{albeverio} for the probability
distribution $\Prob_t[\Rsp]$, which can formally be obtained by inversion of
$Z_t[\eta]$. Hence, up to an overall normalization,
\begin{equation}
  \label{genfunc:prob}
  \Prob_t[\Rsp] \propto \int [\d \eta] \; \exp\left( - \imag \int \d^3 x \;
  \Rsp(\vect{x})\eta(\vect{x}) \right) Z_t[\eta] .
\end{equation}
The normalization is not determined by this procedure, but it is irrelevant.
We will fix the $\Rsp$-independent prefactor, which correctly normalizes
the probability distribution,
by requiring $\int \d \fluct \, \Prob(\fluct) =1$ at the end of the calculation.
For this reason, we systematically drop all field-independent prefactors
in the calculation which follows.

\subsection{The probability density on the ensemble}
So far, all our considerations have been exact, and apply for any
quantum field $\R(t,\vect{x})$. For any such field,
Eq. \eref{genfunc:prob} gives the probability density for a spatial
configuration $\Rsp$ at time $t$, and implies that to obtain $\Prob_t[\Rsp]$
we should need to know \emph{all} such functions for all $n$-point
correlations, and at all spatial positions $\vect{x}$.
In practice, some simplifications occur when $\R$ is identified as the
inflationary curvature perturbation.

The most important simplification is the possibility of a perturbative
evaluation.
The dominant mode of the CMB fluctuation is constrained to be Gaussian
to high accuracy, so the corrections to the leading order Gaussian result
cannot be large. Moreover, since we assume
$\R$ is the curvature perturbation which is communicated to CMB
fluctuations, the amplitude of its spectrum is constrained by observation.
Specifically,
in the region of wavenumbers probed by the COBE DMR instrument
\cite{bennett-banday} (and other later CMB experiments such as
BOOMERANG \cite{boomerang} and WMAP \cite{wmap2006-temp}),
the spectrum has amplitude
$\ps^{1/2} \sim 10^{-5}$, whereas the requirement that inflation not
overproduce PBHs generally requires $\ps^{1/2} \lesssim 10^{-3}$ over
the relevant wavenumbers \cite{carr-constraints}.
Each higher-order correlation function is suppressed by an
increasing number of copies of
the spectrum, $\ps(k)$. Provided the amplitude
of $\ps$ is small, it is reasonable to believe that
we are justified in truncating the exponential in
\eref{genfunc:prob} at a
given level in $n$ and working with a perturbation series in $\ps$.
The relevant criterion is the smallness of the spectrum, rather than the
validity of slow-roll.

However, this simple approach is too na\"{\i}ve, because the integrals over
$\eta$ eventually make any given term in the series large, and invalidate
simple perturbative arguments based on power counting in $\ps$. The perturbation
series can only be justified \emph{a posteriori}, a point to which we will
return in Section~\ref{sec:nongaussian}.

We work to first-order in the three-point correlation,
\begin{equation}
  \label{invert:a}
  \Prob_t[\Rsp] \propto \int [\d \eta] \;
  \correction_t[\eta] \pregauss_t[\eta;\Rsp] ,
\end{equation}
where $\correction[\eta]$ and $\pregauss[\eta;\Rsp]$ are defined by
\begin{equation}
  \fl
  \correction_t[\eta] =
  \left(1 - \frac{\imag}{6}
  \int \frac{\d^3 k_1 \, \d^3 k_2 \, \d^3 k_3}{(2\pi)^9} \;
  \eta(\vect{k}_1) \eta(\vect{k}_2) \eta(\vect{k}_3)
  \langle \R(t,\vect{k}_1) \R(t,\vect{k}_2) \R(t,\vect{k}_3) \rangle \right) ,
\end{equation}
and
\begin{equation}
  \fl
  \pregauss_t[\eta;\Rsp] =
  \exp\left(- \int \frac{\d^3 k_1 \, \d^3 k_2}{(2\pi)^6} \;
  \frac{\eta(\vect{k}_1) \eta(\vect{k}_2)}{2}
  \langle \R(t,\vect{k}_1) \R(t,\vect{k}_2) \rangle - \imag \int
  \frac{\d^3 k}{(2\pi)^3} \;
  \eta(\vect{k}) \Rsp(\vect{k}) \right) .
\end{equation}
In this expression, $\pregauss$ will give rise to the Gaussian part of the
probability distribution, and $\correction$ is of the form $1$ plus a
correction. This correction is small when the perturbative analysis is valid.
Higher-order perturbative corrections in $\ps$ can be accommodated
if desired by
retaining higher-order terms in the power-series expansion of the
exponential in \eref{genfunc:prob}. Therefore our method is not restricted
to corrections arising from non-Gaussianities described by three-point
correlations only, but can account for non-Gaussianities which enter at
any order in the correlations of $\R$, limited only by the computational
complexity. However,
in this paper, we work only with the three-point non-Gaussianity.

We now complete the square in $\pregauss_t[\eta;\Rsp]$ and make
the finite field redefinition
\begin{equation}
  \eta(\vect{k}) \mapsto \hat{\eta}(\vect{k}) = \eta(\vect{k}) + (2\pi)^3
  \imag \frac{\Rsp(\vect{k})}{\langle \R(t,\vect{k}) \R(t,-\vect{k}) \rangle'} ,
\end{equation}
where the prime $'$ attached to $\langle \R(t,\vect{k}) \R(t,-\vect{k})
\rangle'$ indicates that the momentum-conservation $\delta$-function is omitted.
The measure $[\d\eta]$ is formally invariant under this
shift, giving $\int [\d \eta] = \int [\d \hat{\eta}]$,
whereas $\pregauss_t[\eta;\Rsp]$ can be split into an $\Rsp$-dependent piece,
which we call $\Gauss_t[\Rsp]$,
and a piece that depends only on $\hat{\eta}$ but not $\Rsp$,
\begin{equation}
  \pregauss_t[\eta;\Rsp] \mapsto
  \Gauss_t[\Rsp] \exp \left( - \frac{1}{2} \int \frac{\d^3 k_1 \, \d^3 k_2}
  {(2\pi)^6} \; \hat{\eta}(\vect{k}_1) \hat{\eta}(\vect{k}_2)
  \langle \R(t,\vect{k}_1) \R(t,\vect{k}_2) \rangle \right) ,
\end{equation}
where $\Gauss_t[\Rsp]$ is a Gaussian in $\Rsp$,
\begin{equation}
  \fl
  \Gauss_t[\Rsp] = \exp \left( - \frac{1}{2} \int \d^3 k_1 \, \d^3 k_2 \;
  \langle \R(t,\vect{k}_1) \R(t,\vect{k}_2) \rangle
  \frac{\Rsp(\vect{k}_1) \Rsp(\vect{k}_2)}{\prod_i
  \langle \R(t,\vect{k}_i) \R(t,-\vect{k}_i) \rangle'} \right) .
\end{equation}
Eq. \eref{invert:a} for the probability density becomes
\begin{equation}
  \fl\label{invert:b}
  \Prob_t[\Rsp] \propto \Gauss_t[\Rsp] \int [\d \hat{\eta}] \;
  \correction_t
  \exp \left( - \frac{1}{2} \int \frac{\d^3 k_1 \, \d^3 k_2}
  {(2\pi)^6} \; \hat{\eta}(\vect{k}_1) \hat{\eta}(\vect{k}_2)
  \langle \R(t,\vect{k}_1) \R(t,\vect{k}_2) \rangle \right) ,
\end{equation}
One can easily verify that this is the correct expression, since if
we ignore the three-point contribution (thus setting $\correction_t = 1$),
one recovers (after applying a correct normalization)
\begin{equation}
  \int [\d\Rsp] \; \Rsp(\vect{k}_1) \Rsp(\vect{k}_2)
  \Gauss_t[\Rsp] = \langle \R(t,\vect{k}_1) \R(t,\vect{k}_2) \rangle.
\end{equation}
The remaining issue is to carry out the $\hat{\eta}$ integrations
in $\correction_t$. The only terms which contribute are those
containing an even power of $\hat{\eta}$, since any odd function
integrated against $\e{-\hat{\eta}^2}$ vanishes identically.
In the expansion of $\prod_i \eta(\vect{k}_i)$ in terms of $\hat{\eta}$,
there are two such terms: one which is quadratic in $\hat{\eta}$,
and one which is independent of $\hat{\eta}$. These are
accompanied by linear and cubic terms which do not contribute to
$\Prob_t[\Rsp]$. For any symmetric kernel $\mathsf{K}$ and vectors
$\vect{p}$, $\vect{q} \in \reals^m$, one has the general results
\begin{eqnarray}
  \fl
  \int [\d f] \; \exp \left( - \frac{1}{2} \int \d^m x \, \d^m y \;
  f(\vect{x}) f(\vect{y}) \mathsf{K}(\vect{x},\vect{y}) \right) =
  \left( \det \mathsf{K} \right)^{-1/2} , \\
  \fl
  \int [\d f] \; f(\vect{p}) f(\vect{q}) \exp \left( - \frac{1}{2}
  \int \d^m x \, \d^m y \; f(\vect{x}) f(\vect{y}) \mathsf
  {K}(\vect{x},\vect{y}) \right) =
  \mathsf{K}^{-1}(\vect{p},\vect{q}) \left( \det \mathsf{K}
  \right)^{-1/2} .
\end{eqnarray}
These rules allow us to evaluate the $\hat{\eta}$ integrals in
Eq. \eref{invert:b}, giving
\begin{equation}
  \label{invert:c}
  \Prob_t[\Rsp] \propto \Gauss_t[\Rsp] \left(1 + \correction_t^{(0)}[\Rsp]
  + \correction_t^{(2)}[\Rsp] \right) ,
\end{equation}
where $\correction^{(0)}$ and $\correction^{(2)}$ are defined by
\begin{eqnarray}
  \label{invert:d}\fl
  \correction_t^{(0)}[\Rsp]
  = - \frac{1}{6} \int \d^3 k_1 \, \d^3 k_2 \, \d^3 k_3 \;
  \langle \R(t,\vect{k}_1) \R(t,\vect{k}_2) \R(t,\vect{k}_3) \rangle
  \frac{\Rsp(\vect{k}_1) \Rsp(\vect{k}_2) \Rsp(\vect{k}_3)}
  {\prod_i\langle \R(t,\vect{k}_i) \R(t,-\vect{k}_i) \rangle'}, \\
  \nonumber\fl
  \correction_t^{(2)}[\Rsp]
  = \frac{1}{6} \int \d^3 k_1 \, \d^3 k_2 \, \d^3 k_3 \;
  \langle \R(t,\vect{k}_1) \R(t,\vect{k}_2) \R(t,\vect{k}_3) \rangle
  \frac{\Rsp(\vect{k}_1) \diracd(\vect{k}_2 + \vect{k}_3)}
  {\prod_{i \neq 3}\langle \R(t,\vect{k}_i) \R(t,-\vect{k}_i) \rangle'}
  \\ \label{invert:e} \mbox{} + \mbox{combinations} ,
\end{eqnarray}
where in $\correction_t^{(2)}$ we include the possible combinations of
the labels $\{ 1, 2, 3 \}$ which give rise to distinct integrands.

In fact, $\correction_t^{(2)}$ is negligible.
This happens because the three-point function
contains a momentum-conservation $\diracd$-function, $\diracd(\vect{k}_1 +
\vect{k}_2 + \vect{k}_3)$, which requires that the vectors $\vect{k}_i$
sum to a triangle in momentum space. [For this reason, it
is often known as the ``triangle condition'', and
we will usually abbreviate it schematically as
$\diracd(\triangle)$.] In combination with the $\diracd$-%
function $\diracd(\vect{k}_2 + \vect{k}_3)$, the effect is to constrain
two of the momenta (in this example $\vect{k}_2$ and $\vect{k}_3$) to be
equal and opposite, and the other momentum (in this example,
$\vect{k}_1$) to be zero.
This is the extreme \emph{squeezed} limit \cite{maldacena-nongaussian,
creminelli-zaldarriaga,allen-gupta},
in which the bispectrum reduces to the power spectrum evaluated on a perturbed
background, which is sourced by the zero-momentum mode. Written explicitly,
$\correction_t^{(2)}$ behaves like
\begin{equation}
  \label{invert:esti}
  \correction_t^{(2)}[\Rsp]
  \simeq \frac{1}{6} \int \frac{\d^3 k_1 \, \d^3 k_2}{(2\pi)^3}
  \Asq \Rsp(\vect{k}_1)
  \diracd(\vect{k}_1) + \mbox{combinations} ,
\end{equation}
where we have written $\lim_{k_1 \rightarrow 0} \A = \Asq k_2^3$, and
$\Asq$ is a known finite quantity.
In particular, Eq.~\eref{invert:esti} vanishes,
provided $\Rsp(\vect{k})$ approaches zero as $k \rightarrow 0$.
This condition is typically satisfied, since by construction $\Rsp(\vect{k})$
should not contain a zero mode. Indeed, any zero mode, if present, would
constitute part of the zero-momentum background, and not a part of
the perturbation $\Rsp$.
Accordingly, Eqs. \eref{invert:c}--\eref{invert:d} with
$\correction_t^{(2)}=0$ give $\Prob_t[\Rsp]$ explicitly in terms of
the two- and three-point correlation functions.

\subsection{The smoothed curvature perturbation}
The probability density $\Prob_t[\Rsp] \propto (1 + \correction_t^{(0)}
[\Rsp])\Gauss_t[\Rsp]$ that we have derived relates to the microphysical
field $\R(t,\vect{x})$ which appeared in the quantum field theory Lagrangian.
A given $\vect{k}$-mode
of this field begins in the vacuum state at $t \rightarrow -\infty$.
At early times, the mode is far inside the horizon ($k \gg aH$).
In this (``subhorizon'') r\'{e}gime, the $\vect{k}$-mode
cannot explore the curvature
of spacetime and is immune to the fact that it is living in a de Sitter
universe. It behaves like a Minkowski space oscillator.
At late times, the mode is far outside the horizon
($k \ll aH$). In this (``superhorizon'') r\'{e}gime, the $\vect{k}$-mode
asymptotes to a constant amplitude, provided that
only one field is dynamically relevant during inflation
\cite{lyth-malik,wands-malik}.%
\footnote{Where multiple fields are present, there will typically be
an isocurvature perturbation between them: hypersurfaces of constant
pressure and density will not coincide. Under these circumstances
\cite{wands-malik}, $\R$ will evolve. We do not consider the evolving case
in this paper, but rather restrict our attention to the single-field
case where the superhorizon behaviour of $\R$ is simple.}
If we restrict attention to tree-level diagrams, then under reasonable
conditions
the integrals which define the expectation values of $\R$
are typically \cite{weinberg-corrl,weinberg-corrl-a} dominated
by the intermediate (``horizon crossing'') r\'{e}gime,
where $\R(\vect{k})$ is exiting the
horizon ($k \sim aH$).
As a result, the correlation functions
generally depend only on the Hubble and slow-roll parameters around the
time of horizon exit.

The simple superhorizon behaviour of $\R$ means that we can treat the power
spectrum as constant outside the horizon.
As has been described,
its value depends only on the Hubble parameter and the
slow-roll parameters around the
time that the mode corresponding to $k$ exited the horizon.
For this reason, the time $t$ at which we evaluate the wavefunctional
$\Psi_t[\Rsp]$, the generating functional $Z_t[\eta]$ and the
probability distribution $\Prob_t[\Rsp]$ is irrelevant, provided it is taken
to be late enough that the curvature perturbation on interesting cosmological
scales has already been generated and settled down to its final value.
Indeed, we have implicitly been assuming that $t$ is comoving time, so that
observers on slices of constant $t$ see no net momentum flux. Because $\R$
is a gauge invariant, and is constant outside the horizon, our formalism is
independent of how we choose to label the spatial slices.
The evolution of $\R$ outside the horizon is the principal obstacle
which would be involved in extending our analysis to
the multiple-field scenario.

When calculating the statistics of density fluctuations on some given
lengthscale $2\pi/\kmax$, one should smooth the perturbation field over
wavenumbers larger than $\kmax$.
To take account of this, we introduce a smoothed field $\Rsm$ which is related
to $\Rsp$ via the rule $\Rsm(\vect{k}) = \window(k,\kmax)
\Rsp(\vect{k})$, where $\window$ is some window function. The probabilities
we wish to calculate and compare to the real universe relate to $\Rsm$
rather than $\Rsp$.
The exact choice of filter $\window$ is mostly arbitrary. For the
purpose of analytical calculations, it is simplest
to pick a sharp cutoff in $\vect{k}$-space,
which removes all modes with $k > \kmax$.
However, this choice has the disadvantage that it is non-local
and oscillatory in real space, which makes physical interpretations
difficult.
The most common alternative choices, which do not suffer from such drawbacks,
are: (i) a Gaussian, or: (ii)
the so-called ``top hat,'' which is a sharp cutoff in real space.
We allow for a completely general choice of $\cont{0}$ function $\window$,
subject to the restriction that $\window \neq 0$ except at $k = \infty$
and possibly at an isolated set of points elsewhere. This restriction is made
so that there is a one-to-one relationship between $\Rsm$ and $\Rsp$. If this
were not the case, it would be necessary to coarse-grain over classes of
microphysical fields $\Rsp$ which would give rise to the same smoothed
field $\Rsm$.

In addition to this smoothing procedure,
the path integral must be regulated before carrying
out the calculation in the next section. This is achieved by artificially
compactifying momentum space, so that the range of available wavenumbers
is restricted to $k < \Lambda$, where $\Lambda$ is an auxiliary hard cutoff.%
\footnote{Note that this procedure does not have anything to do with
the regularization of ultraviolet divergences.
Such divergences are connected to the appearance of loop
graphs, which we ignore, and in any case are subdominant
\cite{weinberg-corrl,weinberg-corrl-a}.}
At the end of the calculation one takes $\Lambda \rightarrow \infty$.
Some care is necessary in carrying out this compactification.
We set $\Rsm = 0$ for $k > \Lambda$. In order to maintain continuity
at $k = \Lambda$, we introduce a 1-parameter family of
functions $\window_\Lambda$. These functions are supposed to
satisfy the matching condition $\lim_{\Lambda \rightarrow \infty}
\window_\Lambda(k) = \window(k)$, and are subject to the restriction
$\window_\Lambda(\Lambda) = 0$. (These conditions could perhaps be relaxed.)
The relationship between $\Rsp$ and $\Rsm$ becomes
\begin{equation}
  \label{filter:filter}
  \Rsm(\vect{k})
  = \theta(\Lambda-k) \window_\Lambda(k;\kmax) \Rsp(\vect{k}) .
\end{equation}
where $\theta(x) = \int_{-\infty}^x \diracd(z) \, \d z$
is the Heaviside function. To minimise unnecessary
clutter in equations, we frequently suppress the $\Lambda$ and $\kmax$
dependence in $\window$, writing only $\window(k)$ with the smoothing scale
$\kmax$ and hard cutoff $\Lambda$ left implicit.
Both of the standard window functions approach zero as
$k \rightarrow \infty$, and are compatible
with \eref{filter:filter} in the $\Lambda \rightarrow \infty$ limit.
In this limit, the final result is independent of the exact choice of family
$\window_\Lambda(k,\kmax)$.

The probability in which we are interested is that of observing
a given filtered field $\Rsm$. One can express this via the rule
(see also \cite{matarrese-verde,taylor-watts})
\begin{equation}
  \label{filter:prob}
  \Probsm_t[\Rsm]
   = \int [\d \Rsp] \; \Prob_t[\Rsp] \diracd[\Rsm = \theta(\Lambda-k)
   \window \Rsp] .
\end{equation}

\section{Harmonic decomposition of the curvature perturbation}
\label{sec:harmoniccurv}
In the previous section, we obtained the probability density for
a given smoothed spatial configuration of the
curvature perturbation. Given this probability density, the probability
$\Prob(\zeta)$ that the configuration exhibits some characteristic $\zeta$,
such as fluctuations of size $\fluct$ or a `fluctuation spectrum'
(to be defined later) of the form $\spect(k)$, is
formally obtained integrating over all configurations of $\Rsm$
which exhibit the criteria which define $\zeta$
(cf. Ref. \cite{matarrese-verde}).
We give a precise specification of these criteria in
Section~\ref{sec:condition} below. Before doing so, however,
we exploit the compactification
of momentum space introduced in \eref{filter:filter} to define 
a complete set of partial waves.
The smoothed field $\Rsm$ can be written as a superposition of these
partial waves with arbitrary coefficients.
Moreover, the path integral measure can
formally be written as a product of conventional integrals over these
coefficients \cite{hawking-zeta}.

In this section we assemble the necessary
formulae for the partial-wave decomposition.
In particular, we shall require
Eq.~\eref{harmonic:expand} for the decomposition of $\Rsm$,
Eq.~\eref{harmonic:measure} for the path integral measure,
and Eq.~\eref{strength:condition}, which gives a precise specification
of the characteristics $\fluct$ and $\spect(k)$.

\subsection{Harmonic expansion of $\Rsm$}
\label{sec:harmonic}
We expand $\Rsm(\vect{k})$ in harmonics on the unit sphere and along the
radial $k = |\vect{k}|$ direction,
\begin{equation}
  \label{harmonic:expand}
  \Rsm(\vect{k}) = \sum_{\ell = 0}^\infty \sum_{m = - \ell}^{\ell}
  \sum_{n = 1}^\infty \Rsm^m_{\ell|n} Y_{\ell m}(\theta,\phi)
  \psi_n(k) .
\end{equation}
The $Y_{\ell m}(\theta,\phi)$ are the standard spherical harmonics
on the unit 2-sphere (see, \emph{eg}., Ref. \cite{varshalovich}), whereas
the $\psi_n(k)$ are any complete, orthogonal set of functions on the interval
$[0,\Lambda]$. These harmonics should satisfy the following conditions:%
\footnote{When expanding functions on $\reals^3$ in terms of polar coordinates,
a more familiar expansion is in terms of the spherical waves $Z_{\ell m|k}
\propto j_{\ell}(kr)
Y_{\ell m}(\theta,\phi)$, where $j_{\ell}$ is a spherical Bessel function.
These waves are eigenfunctions of the Laplacian in polar coordinates, viz,
$\nabla^2 Z_{\ell m|k} = - k^2 Z_{\ell m |k}$. An arbitrary function on
$\reals^3$ can be written in terms of the spherical waves, which is equivalent
to a Fourier expansion. We do not choose the
spherical waves as an appropriate complete, orthogonal set of basis functions
because we do not wish to expand \emph{arbitrary} functions, but rather
functions obeying particular boundary conditions,
specifically, at $k=0$. The spherical waves
for low $\ell$ behave improperly at small $k$ for this purpose.
Moreover, it is not possible to easily impose the boundary condition
$\Rsm(k) \rightarrow 0$ as $k \rightarrow \Lambda$.}
\begin{enumerate}
  \item $\psi_n(k) \rightarrow 0$ smoothly as $k \rightarrow 0$, so that
        power is cut off on very large scales, and the universe
        remains asymptotically FRW with the zero-mode $a(t)$ which was used when
        computing the expectation values $\langle \R \cdots \R \rangle$
        left intact; \label{harmonic:asymp}
  \item $\psi_n(k) \rightarrow 0$ smoothly as $k \rightarrow \Lambda$,
        so that the resulting $\Rsm$ is compatible with
        Eq.~\eref{filter:filter}; \label{harmonic:compact}
  \item $\psi_n(k)$ should have dimension $\dimension{M^{-3}}$, in order that
        Eq. \eref{harmonic:expand} is dimensionally correct;
        and \label{harmonic:dims}
  \item the $\psi_n(k)$ should be orthogonal in the measure
        $\int_0^\Lambda \d k \, k^5 \ps^{-1}(k)\window^{-2}(k)$.
        \label{harmonic:technical}
\end{enumerate}
In addition, there is a constraint on the coefficients $\Rsm^m_{\ell|n}$,
because $\Rsm(\vect{k})$ should be real in configuration space and
therefore must obey the Fourier reality
condition $\Rsm(\vect{k})^\ast = \Rsm(-\vect{k})$, where a star
`$\ast$' denotes complex conjugation. The $\Rsm^m_{\ell|n}$
are generically complex, so it is useful to separate the real and imaginary
parts by writing $\Rsm^m_{\ell|n} = a^m_{\ell|n} + \imag b^m_{\ell|n}$.
The condition that $\Rsm$ is real in configuration space implies
\begin{eqnarray}
  \label{harmonic:reality}
  a^{-m}_{\ell|n} = (-1)^{\ell+m}a^m_{\ell|n} \\
  b^{-m}_{\ell|n} = (-1)^{\ell+m+1}b^m_{\ell|n} .
\end{eqnarray}
These conditions halve the number of independent coefficients,
since the $a$ and $b$ coefficients with strictly
negative $m$ are related to those with
strictly positive $m$, whereas for the $m=0$ modes,
the $b$ coefficients vanish if $\ell$ is even, and the $a$ coefficients
vanish if $\ell$ is odd.

Condition~(\ref{harmonic:asymp}) is made so that the power on large
scales is smoothly cut off. In the absence of this constraint, $\Rsm$ could
develop unbounded fluctuations on extremely large scales which would renormalize
$a(t)$. Therefore, Condition~(\ref{harmonic:asymp})
can be interpreted as a consistency requirement, since the inflationary
two- and three-point functions are calculated using perturbation theory
on an FRW universe with some given $a(t)$, which must be recovered
asymptotically as $|\vect{x}| \rightarrow \infty$.
It will later be necessary to sharpen
this condition to include constraints on the behaviour of $\ps(k)$ near
$k=0$, beyond the weak requirement that $\sigma^2 = \int \ps(k)
\, \d \ln k$ is finite.
Condition~(\ref{harmonic:technical})
is a technical condition made for future convenience. Any other choice of
normalization would work just as well, but this choice is natural
given the $k$-dependence in the Gaussian kernel $\Gaussm[\Rsm]$. Indeed,
with this condition, the Gaussian prefactor in $\Prob
(\fluct)$ will reduce to the exponential of a sum of squares of the
$a^m_{\ell|n}$ and $b^m_{\ell|n}$.
We stress that in virtue of Condition (\ref{harmonic:dims}), the inner product
of two $\psi_n(k)$ in the measure $\int_0^\Lambda \d k \, k^5 \ps^{-1}(k)
\window^{-2}(k)$ is dimensionless.
Condition~(\ref{harmonic:compact}) has less fundamental significance.
It follows from the condition $\window_\Lambda(\Lambda) = 0$ and the
artificial compactification of momentum space. However, as in the usual
Sturm--Liouville theory \cite{morse-feshbach}, the precise
choice of boundary condition is immaterial when the regulator is removed
and $\Lambda \rightarrow \infty$. Condition~(\ref{harmonic:compact})
does not affect the final answer.

To demonstrate the existence of a suitable set of $\psi_n(k)$,
we can adopt the definition
\begin{equation}
  \label{harmonic:psi}
  \psi_n(k) = \frac{\sqrt{2}}{J_{\nu+1}(\alpha_\nu^n)} \frac{\ps(k)
  \window(k)}
  {\Lambda k^2} J_\nu\left( \alpha_\nu^n \frac{k}{\Lambda} \right) ,
\end{equation}
where $J_\nu(z)$ is the Bessel function of order $\nu$, which is regular
at the origin, and
$\alpha_\nu^n$ is its $n$th zero. The order $\nu$ is arbitrary, except that
in order to obey condition (\ref{harmonic:asymp}) above, we must have
$k^{\nu-2} \ps(k) \rightarrow 0$ as $k \rightarrow 0$, assuming that
$\window(k) \rightarrow 1$ as $k \rightarrow 0$ (as is proper for a
volume-normalized window function).
The $\psi_n(k)$ obey the orthonormality condition
\begin{equation}
  \label{harmonic:orthonormal}
  \int_0^\Lambda \d k \; \frac{k^5}{\ps(k) \window^2(k)}
  \psi_n(k) \psi_m(k) = \delta_{mn} ,
\end{equation}
and $\delta_{mn}$ is the Kronecker delta. The completeness relation can be
written
\begin{equation}
  \label{harmonic:complete}
  \diracd(k-k_0)|_{k \in [0,\Lambda]}
  = \frac{k_0^5}{\ps(k_0) \window^2(k_0)} \sum_n \psi_n(k) \psi_n(k_0) ,
\end{equation}
where the range of $k$ is restricted to the compact interval $[0,\Lambda]$.

Although we have given an explicit form for the $\psi_n$ in order to
demonstrate existence, the argument does not depend in detail
on Eq.~\eref{harmonic:psi}. The only important properties are
Eqs.~\eref{harmonic:orthonormal}--\eref{harmonic:complete}, which
follow from Condition~(\ref{harmonic:technical}).

\subsection{The path integral measure}
\label{sec:pathmeasure}
Since any real, $\cont{0}$ function $\Rsm$
obeying the boundary conditions $\Rsm(\vect{k})
\stackrel{k \rightarrow 0}{\longrightarrow} 0$ and
$\Rsm(\vect{k}) \stackrel{k \rightarrow \Lambda}{\longrightarrow} 0$ can
be expanded in the form \eref{harmonic:expand}, one can formally integrate
over all such $\Rsm$ by integrating over the coefficients $\Rsm^m_{\ell|n}$.
This prescription has been widely used for obtaining explicit results from
path integral calculations. (For a textbook treatment, see Ref.
\cite{kleinert}.) In the present case, it must be remembered that one
should include in the integral only those $\Rsm(\vect{x})$ which are real,
and can therefore correspond to a physical curvature perturbation in the
universe. Since the $Y_{\ell m}$ are complex,
this means that instead of integrating unrestrictedly over the
$\Rsm^m_{\ell|n}$, the reality conditions \eref{harmonic:reality} must
be respected. A simple way to achieve this is to integrate only over
those $a^m_{\ell|n}$ or $b^m_{\ell|n}$ with $m \geq 0$. The $m=0$ modes
must be treated specially since the $a$ and $b$ coefficients vanish for
odd and even $\ell$, respectively.

The integral over real $\Rsm$ can now be written
\begin{equation}
  \fl\label{harmonic:measure}
  \int_{\reals} [\d \Rsm] =
  \Bigg[ \prod_{\ell =0}^{\infty} \prod_{m=1}^\infty \prod_{n=1}^{\ell}
  \mu \int_{-\infty}^{\infty} \d a^m_{\ell|n} \int_{-\infty}^\infty
  \d b^m_{\ell|n} \Bigg] \Bigg[ \prod_{\substack{r=0 \cr \even{r}}}
  ^\infty \prod_{s = 1}^\infty
  \tilde{\mu} \int_{-\infty}^\infty \d a^0_{r|s} \int_{-\infty}^\infty
  \d b^0_{r+1|s} \Bigg] ,
\end{equation}
where the subscript $\reals$ on the integral indicates schematically that
only real $\Rsm(\vect{x})$ are included.
The constants $\mu$ and $\tilde{\mu}$ account
for the Jacobian determinant which arises in writing $\int [\d\Rsm]$ in terms
of the harmonic coefficients $\Rsm^m_{\ell|n}$. Their precise form is of no
importance in the present calculation.

As we noticed above, the detailed
form of the measure \eref{harmonic:measure} is not absolutely necessary
for our argument. The important aspect is that each $a$ or $b$ integral
(where $m \geq 0$)
can be carried out independently.
For this purpose it is sufficient that the spectrum of partial waves is
discrete, which follows from the (artificial) compactness of momentum space.
However, although it is necessary to adopt some regulator in order to
write the path integral measure in a concrete form such as
\eref{harmonic:measure}, we expect the answer to be independent of the
specific regulator which is chosen. In the present context, this means that
our final expressions should not depend on $\Lambda$, so that the passage
to the $\Lambda \rightarrow \infty$ limit becomes trivial.

\subsection{The total fluctuation $\fluct$ and the spectrum $\spect(k)$}
\label{sec:condition}
There are at least two useful ways in which one might attempt to measure
the strength of fluctuations in $\Rsm$.
The first is the \emph{total smoothed fluctuation}
at a given point $\vect{x} = \vect{x}_0$. By
a suitable choice of coordinates, we can always arrange that $\vect{x}_0$
is the origin, so the condition is $\Rsm(\vect{0}) = \fluct$.
When $\Rsm$ is smoothed on scales of order the horizon size this gives
a measure of the fluctuation in each Hubble volume, since distances of
less than a horizon size no longer have any meaning.
For example,
Shibata \& Sasaki \cite{shibata-sasaki} have proposed that $\fluct$
defined in this way
represents a useful criterion for the formation of PBHs, with formation
occuring whenever $\fluct$ exceeds a threshold value $\fluct_{\mathrm{th}}$
of order unity \cite{green-liddle}.
This measure of the fluctuation is non-local in momentum space.
Making use of the relation $\int \d\Omega(\theta,\phi) \, Y_{\ell m}(\theta,
\phi) = \sqrt{4\pi} \delta_{\ell,0} \delta_{m,0}$ for the homogeneous mode
of the spherical harmonics, one can characterize $\fluct$ as
\begin{equation}
  \label{strength:nonlocal}
  \int \frac{\d^3 k}{(2\pi)^3} \Rsm(\vect{k}) \e{\imag \vect{k}\cdot\vect{x}}|
  _{\vect{x}=0} = \frac{\sqrt{4\pi}}{(2\pi)^3} \int \d k \; k^2
  \sum_{n=1}^{\infty} a^0_{0|n} \psi_n(k) = \fluct .
\end{equation}

On the other hand, one might be interested in contributions to the
total smoothed fluctuation
in each Hubble volume which arise from features in the spectrum near some
characteristic scale of wavenumber $k$. For this reason, we consider
a second possible measure of fluctuations, which we call the
\emph{fluctuation spectrum}, and which is defined by the condition
$\spect(k) = \d \Rsm(\vect{0}) / \d \ln k$.
(Thus, the total smoothed fluctuation can be obtained by integrating
its spectrum according to the usual rule, viz, $\fluct = \int \spect(k) \, \d
\ln k$.) This condition is local
in $\vect{k}$-space. Differentiating \eref{strength:nonlocal},
one can characterize $\spect(k)$ as
\begin{equation}
  \label{strength:local}
  \spect(k) = \frac{\sqrt{4\pi}}{(2\pi)^3}
  \sum_{n=1}^{\infty} a^0_{0|n} k^3 \psi_n(k) .
\end{equation}
This is a functional constraint.

We will calculate the statistics of both the total
fluctuation $\fluct$ and its spectrum $\spect(k)$. In each
case, the calculation is easily adapted to other observables which are
non-local or local in momentum space, respectively. Indeed, both the
non-local $\fluct$ and the local $\spect(k)$ are members of a large class
of observables, which we can collectively denote $\zeta$, and
which all share nearly-Gaussian statistics. Specifically, Eqs.
\eref{strength:nonlocal} and \eref{strength:local} can be written in a unified
manner in the form
\begin{equation}
  \label{strength:condition}
  \sum_{n=1}^\infty a^0_{0|n} \Sigma_n(k)
  = \frac{(2\pi)^3}{\sqrt{4\pi}} \zeta(k) ,
\end{equation}
where the $\Sigma_n$ are defined by
\begin{equation}
  \Sigma_n = \left\{ \begin{array}{ll}
  \int_0^\Lambda \d k \; k^2 \psi_n(k) & \mbox{total fluctuation},
    \zeta = \fluct; \\
  k^3 \psi_n(k) & \mbox{fluctuation spectrum},
    \zeta = \spect(k) .
  \end{array} \right.
\end{equation}
Note that in the case of $\fluct$, the $\Sigma_n$ are independent of $k$.
Any characteristic which can be put in this form, with a coupling only to the
real zero-modes $a^0_{0|n}$ of $\Rsm$, will necessarily develop
nearly-Gaussian (\emph{i.e.}, weakly non-Gaussian) statistics.
More general choices of characteristic are possible,
which cannot be cast in the form \eref{strength:condition}.
For example, one can consider characteristics which depend non-linearly on
the $a^0_{0|n}$. Such characteristics
will generally lead to strongly non-Gaussian probabilities.
The Gaussianity of the final probability distribution depends on the geometry
of the constraint surface in an analogous way to the decoupling of the
Fadeev-Popov ghost fields in gauge field theory \cite{weinberg}.
These non-Gaussian choices of characteristic can also
be handled by generalizing our technique, but we do not consider them here.

\section{The probability density function for $\fluct$}
\label{sec:probdelta}
We first calculate the probability density for the non-local constraint
$\fluct$, given by Eq.~\eref{strength:nonlocal}.
The expression is
\begin{equation}
  \label{prob:defn}
  \Prob(\fluct) \propto \int_{\reals} [\d \Rsm] \; \Probsm[\Rsm]
  \diracd\left[ \sum_{n=1}^\infty a^0_{0|n} \Sigma_n -
  \frac{(2\pi)^3}{\sqrt{4\pi}} \fluct \right] .
\end{equation}
To obtain this density, one treats $\fluct$ as a collective coordinate
parametrizing part of $\Rsm$. The remaining degrees of freedom, which are
orthogonal to $\fluct$, are denoted $\Rsm^\perp$. Therefore the
functional measure can be broken into
$[\d \Rsm] \propto [\d \Rsm^\perp]\, \d\fluct$. After integrating
the functional density $\Prob[\Rsm] \, [\d \Rsm]$ over
$\Rsm^\perp$, the quantity which is left is the probability density
$\Prob(\fluct)\,\d\fluct$.
In this case, the integration over the orthogonal degrees of freedom
$\Rsm^\perp$ is accomplished via the
$\diracd$-function, which filters out only those members of the ensemble which
satisfy Eq~\eref{strength:nonlocal}.
We emphasize that this is a conventional
$\diracd$-function, not a $\diracd$-functional. There is no need to
take account of a Fadeev--Popov type factor because the
Jacobian associated with
the constraint \eref{strength:condition} is field-independent, in virtue
of the linearity of \eref{strength:nonlocal} in $a^0_{0|n}$.

\subsection{The Gaussian case}
\label{sec:gaussian}
We first give the calculation in the approximation that only the two-point
function is retained. In this approximation, the probability distribution
of $\fluct$ will turn out to be purely Gaussian, which allows us to develop
our method without the distractions introduced by including non-Gaussian
effects.

If all correlation functions of order three and higher are set to zero, then
$\Prob[\Rsm] \propto \Gaussm[\Rsm]$. Using \eref{intro:twopt}, one can write
\begin{eqnarray}
  \fl\nonumber
  \Gaussm[\Rsm] = \exp \Bigg( - \frac{1}{2} \int \d\Omega \int k^2 \, \d k \;
  \frac{k^3}{(2\pi)^3 2\pi^2} \frac{1}{\ps(k)\window^2(k)}
  \\ \mbox{} \times
  \sum_{\ell_1, m_1, n_1} \sum_{\ell_2, m_2, n_2} \Rsm^{m_1}_{\ell_1|n_1}
  \Rsm^{m_2\dag}_{\ell_2|n_2} Y_{\ell_1 m_1}(\theta,\phi)
  Y^\dag_{\ell_2,m_2}(\theta,\phi) \psi_{n_1}(k) \psi_{n_2}(k)
  \Bigg) .
\end{eqnarray}
The harmonics $Y_{\ell m}$ and $\psi_n$ integrate out of this expression
entirely, using the orthonormality relation \eref{harmonic:orthonormal}
and the spherical harmonic completeness relation $\int \d \Omega \,
Y_{\ell_1 m_1} Y^\dag_{\ell_2 m_2} = \delta_{\ell_1 \ell_2}
\delta_{m_1 m_2}$. Moreover, after rewriting the $a$ and $b$ coefficients with
$m<0$ in terms of the $m>0$ coefficients, we obtain
\begin{equation}
  \fl
  \Gauss[\Rsm] = \exp\Bigg( - \frac{1}{2\pi^2 (2\pi)^3}
  \sum_{\ell =0}^\infty \sum_{m=1}^{\ell} \sum_{n=1}^\infty
  |a^m_{\ell|n}|^2 + |b^m_{\ell|n}|^2 -
  \frac{1}{4\pi^2 (2\pi)^3} \sum_{\substack{\ell = 0 \cr \even{\ell}}}
  ^\infty \sum_{n=1}^\infty |a^0_{\ell|n}|^2 + |b^0_{\ell+1|n}|^2
  \Bigg) .
\end{equation}

The $\diracd$-function in \eref{prob:defn} constrains one of the
$a^0_{0|n}$ in terms of $\fluct$ and the other coefficients. It is
possible to evaluate $\Prob(\fluct)$ by integrating out the $\diracd$-function
immediately. This would involve solving the constraint
for $a^0_{0|0}$ (for example)
and replacing it in the integrand with its expression in
terms of the other $a^0_{0|n}$ and $\fluct$.
However, this does not turn out to be a
convenient procedure, for the same reasons that one encounters when
gauge-fixing in field theory.
Instead, we introduce the Fourier representation of the
$\diracd$-function,
\begin{equation}
  \fl \Prob(\fluct) \propto \int_{\reals} [\d \Rsm] \int_{-\infty}^\infty
  \d z \; \Gaussm[\Rsm] \exp \left( \imag z \left[ \sum_{n=1}^\infty a^0_{0|n}
  \Sigma_n - \frac{(2\pi)^3}{\sqrt{4\pi}} \fluct \right] \right) ,
\end{equation}
where the functional measure is understood to be Eq. \eref{harmonic:measure}.
The final answer is obtained by integrating out $z$ together with all of
the $a$ and $b$
coefficients. In order to achieve this, it is necessary to decouple
$a^0_{0|n}$, $z$ and $\fluct$ from each other by successively
completing the square in $a^0_{0|0}$ and $z$. Working with $a^0_{0|0}$ first,
we find
\begin{eqnarray}
  \nonumber
  \fl \exp\left( - \frac{1}{4\pi^2} \frac{1}{(2\pi)^3} \sum_{n=1}^\infty
  |a^0_{0|n}|^2 + \imag z \sum_{n=1}^\infty a^0_{0|n} \Sigma_n \right) \\
  \label{gauss:asquare}
  \lo{=} \exp \left( - \frac{1}{4\pi^2} \frac{1}{(2\pi)^3} \sum_{n=1}^\infty
  (a^0_{0|n} - 2 \pi^2 (2\pi)^3 \imag z \Sigma_n )^2 -
  (2\pi)^3 \pi^2 z^2 \Sigma^2 \right) ,
\end{eqnarray}
where we have introduced a function $\Sigma^2$, defined by $\Sigma^2 = \sum
_{n=1}^\infty \Sigma_n^2$. In the final probability distribution,
$\Sigma^2$ will turn out to be the variance
of $\fluct$. From Eq. \eref{gauss:asquare},
it is clear that making the transformation
$a^0_{0|n} \mapsto a^0_{0|n} + 2\pi^2 (2\pi)^3 \imag z \Sigma_n$ suffices
to decouple $a^0_{0|n}$ from $z$. The measure, Eq. \eref{harmonic:measure},
is formally invariant under this transformation. Exactly the same procedure
can now be applied to $z$ and $\fluct$, giving
\begin{equation}
  \fl \exp \left( - (2\pi)^3 \pi^2 z^2 \Sigma^2 - \frac{(2\pi)^3}{\sqrt{4\pi}}
  \imag \fluct z \right) = \exp \left[ - (2\pi)^3 \pi^2 \Sigma^2 \left(
  z + \frac{\imag \fluct}{2 \pi^2 \sqrt{4\pi} \Sigma^2} \right)^2
  - \frac{\fluct^2}{2\Sigma^2} \right] .
\end{equation}
As before, the finite shift $z \mapsto z - \imag \fluct / 2\pi^2 \sqrt{4\pi}
\Sigma^2$ leaves the measure intact and decouples $z$ and $\fluct$. The
$a$, $b$ and $z$ integrals can be done independently, but since they do not
involve $\fluct$ they contribute only an irrelevant normalization to
$\Prob(\fluct)$. Thus, we obtain Gaussian statistics for $\fluct$,
\begin{equation}
  \label{gauss:gauss}
  \Prob(\fluct) \propto \exp \left( - \frac{\fluct^2}{2\Sigma^2} \right) .
\end{equation}

It remains to evaluate the variance $\Sigma^2$.
In the present case, we have $\Sigma_n = \int_0^\Lambda
\d k \, k^2 \psi_n(k)$. From the completeness relation
Eq. \eref{harmonic:complete}, it follows that
\begin{equation}
  \sum_n k_0^2 \psi_n(k_0) k^2 \psi_n(k) = \frac{k^2 \ps(k_0)
  \window^2(k_0)} {k_0^3} \diracd(k-k_0) .
\end{equation}
$\Sigma^2$ is now obtained by integrating term-by-term under the summation.
The result coincides with the \emph{smoothed} conventional variance
(compare Eq.~\eref{intro:variance}),
\begin{equation}
  \label{total:variance}
  \Sigma^2_{\Lambda}(\kmax)
  = \int_0^\Lambda \d \ln k \; \window^2(k;\kmax) \ps(k) .
\end{equation}
Thus, as expected, Eq. \eref{gauss:gauss} reproduces the
Gaussian distribution \eref{intro:gaussian}
which was derived on the basis of the central limit theorem, with the
proviso that the parameters (such as $\Sigma^2$) describing the distribution
of $\fluct$ are associated with the smoothed field $\Rsm$ rather than the
microphysical field $\Rsp$. Note that $\Sigma^2$ is therefore implicitly
a function of scale, with the scale dependence
entering through the window function.

In particular,
it was only necessary to use the completeness relation to obtain this
result, which follows from Condition~(\ref{harmonic:technical}) in
Section~\ref{sec:harmonic}.

\subsection{The non-Gaussian case}
\label{sec:nongaussian}
The non-Gaussian case is a reasonably straightforward extension of the
calculation described in the preceding section, with the term
$\correction^{(0)}$ in Eq. \eref{invert:c} (which was dropped in the
previous section) included. However,
some calculations become algebraically long, and there are subtleties
connected to the appearance of the bispectrum.

The inclusion of $\correction^{(0)}$ corrects the pure Gaussian statistics
by a quantity
proportional to the three-point function, $\langle \R \R \R \rangle$,
which is given in Eq.~\eref{intro:threept}.
This correction is written in terms of the
representative spectrum $\bar{\ps}^2$, which describes when the slow-roll
prefactor, given by the amplitude of the spectrum, should be evaluated
\cite{maldacena-nongaussian}.
For modes which cross the horizon
almost simultaneously, $k_1 \sim k_2 \sim k_3$, this prefactor should be
$\bar{\ps}^2 = \ps(k)^2$, where $k$ is the common magnitude of the
$k_i$. In the alternative case where one
$\vect{k}$-mode crosses appreciably before the other two, $\bar{\ps}^2$
should be roughly given by
\begin{equation}
  \label{intro:prefactor}
  \bar{\ps}^2 = \ps(\max k_i) \ps(\min k_i) .
\end{equation}
Since the difference between this expression and $\ps(k)^2$ when all $k$
are of the same magnitude is very small, it is reasonable to adopt
Eq. \eref{intro:prefactor} as our definition of $\bar{\ps}^2$. We stress
that this prescription relies on the conservation of $\R$ outside the
horizon \cite{allen-gupta}, and
therefore would become more complicated if extended
to a multiple field scenario.

With this parametrization, the probability measure on the ensemble is
obtained by combining \eref{intro:twopt}, \eref{invert:c},
\eref{invert:d} and \eref{intro:threept},
\begin{equation}
  \fl
  \Probsm[\Rsm] \propto \Gaussm[\Rsm] \left( 1 - \frac{1}{6}
  \int \frac{\d^3 k_1 \, \d^3 k_2 \, \d^3 k_3}{(2\pi)^6 2\pi^2}
  \diracd(\triangle) \frac{\bar{\ps}^2 \A}{\prod_i \ps(k_i)}
  \frac{\Rsm(\vect{k}_1) \Rsm(\vect{k}_2) \Rsm(\vect{k}_3)}
  {\window(k_1) \window(k_2) \window(k_3)} \right) .
\end{equation}
This expression should be integrated with the constraint
\eref{strength:nonlocal} and measure \eref{harmonic:measure}
to obtain the probability $\Prob(\fluct)$.
At first this appears to lead to an undesirable consequence, since the
integral of any odd function of $\Rsm$ against $\Gaussm[\Rsm]$ must be zero.
It may therefore seem as if the non-Gaussian corrections we are trying
to obtain will evaluate to zero. This would certainly be correct if the
integral were unconstrained. However, the presence of the constraint
$\diracd$-function
means that the shifts of $a^0_{0|n}$ and $z$ which are
necessary to decouple the integration variables give rise to a non-vanishing
correction.

The finite shift necessary to decouple $a^0_{0|n}$ and $z$ is not changed
by the presence of non-Gaussian corrections, since it only depends on the
argument of the exponential term. This is the same in the Gaussian and
non-Gaussian cases. After making this shift, which again leaves the
measure invariant, the integration becomes
\begin{equation}
  \fl\label{nongauss:a}
  \Prob(\fluct) \propto \int_{\reals} [\d \Rsm] \int_{-\infty}^\infty
  \d z \; \Gaussm[\Rsm] \exp\left( - (2\pi)^3 \pi^2 \Sigma^2 z^2 -
  \frac{(2\pi)^3}{\sqrt{4\pi}} \imag z \fluct \right)
  (1 - J_0 - J_2) ,
\end{equation}
where $J_0$ is given by
\begin{eqnarray}
  \fl\nonumber
  \int \d^3 k_1 \, \d^3 k_2 \, \d^3 k_3 \frac{2\pi^4(2\pi)^3}{3
  (4\pi)^{3/2}}
  \diracd(\triangle) \frac{\bar{\ps}^2 \A}{\prod_i \ps(k_i)}
  \sum_{n_1, n_2, n_3} \imag^3 z^3
  \Sigma_{n_1} \Sigma_{n_2} \Sigma_{n_3}
  \frac{\psi_{n_1}(k_1) \psi_{n_2}(k_2) \psi_{n_3}(k_3)}
  {\window(k_1) \window(k_2) \window(k_3)} ,
\end{eqnarray}
and $J_2$ is
\begin{eqnarray}
  \fl\nonumber \Bigg[
  \int \frac{\d^3 k_1 \, \d^3 k_2 \, \d^3 k_3}{6(2\pi)^3 \sqrt{4\pi}}
  \diracd(\triangle) \frac{\bar{\ps}^2 \A}{\prod_i \ps(k_i)}
  \sum_{n_1} \sum_{\ell_2, m_2, n_2} \sum_{\ell_3, m_3, n_2}
  \\ \nonumber \mbox{} \times
  \imag z \Sigma_{n_1}
  \frac{\psi_{n_1}(k_1)}{\window(k_1)}
  \Rsm^{m_2}_{\ell_2|n_2} \Rsm^{m_3}_{\ell_3|n_3}
  Y_{\ell_2 m_2}(\theta_2,\phi_2)Y_{\ell_3 m_3}(\theta_3,\phi_3)
  \frac{\psi_{n_2}(k_2) \psi_{n_3}(k_3)}{\window(k_2)\window(k_3)} \Bigg] \\
  \fl \hspace{1cm} \mbox{} + \swaplabel{1}{2} + \swaplabel{1}{3} .
\end{eqnarray}
The symbol $\swaplabel{1}{2}$ represents the expression in square
brackets $[ \cdots ]$ with the label $1$
exchanged with the label $2$, and similarly for $\swaplabel{1}{3}$.
The range of the $m_2$, $m_3$ summations is from $-\ell_2$ to $\ell_2$
and $-\ell_3$ to $\ell_3$, respectively. In addition, the shift
of $a^0_{0|n}$ generates other terms linear and cubic in the
$\Rsm^{m}_{\ell|n}$,
but these terms do not contribute to $\Prob(\fluct)$ and we have omitted them
from \eref{nongauss:a}.

After shifting $z$ to decouple $z$ and $\fluct$, the integrals $J_0$ and $J_2$
develop terms proportional to $z^0$, $z$, $z^2$ and $z^3$. Of these, only
the $z^0$ and $z^2$ survive the final $z$ integration. Correspondingly, we
suppress terms linear and cubic in $z$ from the following expressions.
The integral $J_0$ becomes
\begin{eqnarray}
  \nonumber
  \fl J_0 = \int \d^3 k_1 \, \d^3 k_2 \, \d^3 k_3 \;
  \frac{\pi^2(2\pi)^3}{3(4\pi)^2} \left( \frac{1}{16\pi^5}
  \frac{\fluct^3}{\Sigma^6} - 3 \frac{z^2 \fluct}{\Sigma^2} \right)
  \diracd(\triangle) \frac{\bar{\ps}^2 \A}{\prod_i \ps(k_i)}
  \\ \mbox{} \times
  \sum_{n_1, n_2, n_3} \Sigma_{n_1} \Sigma_{n_2} \Sigma_{n_3}
  \frac{\psi_{n_1}(k_1) \psi_{n_2}(k_2) \psi_{n_3}(k_3)}
  {\window(k_1) \window(k_2) \window(k_3)} ,
\end{eqnarray}
whereas $J_2$ simplifies to
\begin{eqnarray}
  \fl\nonumber
  J_2 = \Bigg[ \int \frac{\d^3 k_1 \, \d^3 k_2 \, \d^3 k_3}{48\pi^3(2\pi)^3}
  \frac{\fluct}{\Sigma^2} \sum_{n_1} \sum_{\ell_2, m_2, n_2}
  \sum_{\ell_3, m_3, n_3} \\ \nonumber \mbox{} \times
  \Sigma_{n_1} \frac{\psi_{n_1}(k_1)}{\window(k_1)} \Rsp^{m_2}_{\ell_2|n_2}
  \Rsp^{m_3}_{\ell_3|n_3} Y_{\ell_2 m_2}(\theta_2,\phi_2)
  Y_{\ell_3 m_3}(\theta_3,\phi_3) \frac{\psi_{n_2}(k_2) \psi_{n_3}(k_3)}
  {\window(k_2)\window(k_3)} \Bigg]
  \\
  \fl \hspace{1cm} \mbox{} + \swaplabel{1}{2} + \swaplabel{1}{3} ,
\end{eqnarray}
and the $m$ summations are still over the entire range,
$-\ell_2 \leq m_2 \leq \ell_2$ (and similarly for $m_3$). Thus $J_0$ contains
corrections proportional to $\fluct$ and $\fluct^3$, whereas $J_2$
only contains corrections proportional to $\fluct$.

The $a$, $b$ and $z$ integrations can now be performed, after the integrand
has been written entirely in terms of the $a^m_{\ell|n}$ and $b^m_{\ell|n}$
with $m \geq 0$. There are no $a$ or $b$ integrations in $J_0$.
In $J_2$, there are no $z$ integrations, but
the $a$ and $b$ integrations involved in the product
$\Rsm^{m_2}_{\ell_2|n_2} \Rsm^{m_3}_{\ell_3|n_3}$ fix $\ell_2 = \ell_3$,
$m_2 = m_3$ and $n_2 = n_3$. One then uses the spherical harmonic completeness
relation,
\begin{equation}
  \fl
  \sum_{\ell = 0}^\infty \sum_{m = -\ell}^\ell
  Y_{\ell m}(\theta_1,\phi_1) Y_{\ell m}^\dag(\theta_2,\phi_2) =
  \diracd(\phi_1 - \phi_2) \diracd(\cos \theta_1 - \cos\theta_2)
\end{equation}
and the equivalent relationship for the $\psi$-harmonics, Eq.
\eref{harmonic:complete}, to obtain
\begin{eqnarray}
  \fl\nonumber
  J_2 = \Bigg[ \int \frac{\d^3 k_1 \, \d^3 k_2 \d^3 k_3}{24\pi} \frac{\fluct}
  {\Sigma^2} \diracd(\triangle) \frac{\bar{\ps}^2 \A}{\prod_i \ps(k_i)
  \window(k_i)}
  \frac{\ps(k_2) \window^2(k_2)}{k_2^3}
  \sum_n \Sigma_n \psi_n(k_1)
  \diracd(\vect{k}_2 + \vect{k}_3) \Bigg] \\ \fl \hspace{1cm} \mbox{} +
  \swaplabel{1}{2} + \swaplabel{1}{3} .
\end{eqnarray}
The terms with 1 exchanged with 2 and 3 do not generate quantitatively
different integrands and can be absorbed into an overall factor of 3.

$J_0$ involves only $z$ integrations. It can be written
\begin{eqnarray}
  \fl\nonumber
  J_0 = \int \frac{\d^3 k_1 \, \d^3 k_2 \, \d^3 k_3}{96\pi^2} \left(
  \frac{\fluct^3}{\Sigma^6} - 3 \frac{\fluct}{\Sigma^4} \right)
  \diracd(\triangle) \frac{\bar{\ps}^2 \A}{\prod_i \ps(k_i)\window(k_i)}
  \\ \mbox{} \times
  \sum_{n_1, n_2, n_3} \Sigma_{n_1} \Sigma_{n_2} \Sigma_{n_3}
  \psi_{n_1}(k_1) \psi_{n_2}(k_2) \psi_{n_3}(k_3) .
\end{eqnarray}

To simplify these expressions further, it is necessary to obtain
the value of the sum $\sum_{n=1}^\infty \Sigma_n \psi_n(k)$.

Reasoning as before from the completeness relation
Eq.~\eref{harmonic:complete}, it follows that
\begin{equation}
  \sum_{n=1}^\infty \Sigma_n \psi_n(k) = \frac{\ps(k)\window^2(k)}{k^3} .
\end{equation}
From this, it is straightforward to show that $J_0$ behaves like
\begin{equation}
  J_0 = \int \frac{\d^3 k_1 \, \d^3 k_2 \, \d^3 k_3}
  {96\pi^2\prod_i k_i^3 \window^{-1}(k_i)}
  \diracd(\triangle) \bar{\ps}^2 \A \left(
  \frac{\fluct^3}{\Sigma^6} - 3 \frac{\fluct}{\Sigma^4} \right) ,
\end{equation}
where $\Sigma^2$ is the smoothed variance, Eq. \eref{total:variance}.
On the other hand $J_2$ becomes
\begin{equation}
  J_2 = \int \frac{\d^3 k_1 \, \d^3 k_2 \, \d^3 k_3}{24\pi/3}
  \frac{\fluct}{\Sigma^2} \diracd(\triangle) \frac{\bar{\ps}^2}
  {\ps(k_2)}
  \window(k_1) \A \frac{\diracd(\vect{k}_2 + \vect{k}_3)}{k_1^3 k_2^3} .
\end{equation}
After integrating out $\vect{k}_3$ and the angular part of $\vect{k}_1$
and $\vect{k}_2$, this is the same as
\begin{equation}
  \fl
  J_2 = 2\pi \int \d k_2 \; k_2^2 \int \d k_1 \; \diracd(k_1)
  \frac{\fluct}{\Sigma^2} \window(k_1) \frac{1}{k_2^3 \ps(k_2)}
  \lim_{k_1 \rightarrow 0} \A \frac{\ps(k_1)}{k_1^3} ,
\end{equation}
where we have used the fact that $k_1$ is constrained to zero by the
$\diracd$-function to evaluate the bispectrum $\A$ in the `squeezed'
limit where one of the momenta goes to zero \cite{maldacena-nongaussian,
allen-gupta,creminelli-zaldarriaga}. In this limit, $\min k_i = k_1$
and $\max k_i = k_2 = k_3$, so it is possible to expand $\bar{\ps}^2$
unambiguously. Moreover, $\lim_{k_1 \rightarrow 0} \A = \Asq k_2^3$
is proportional to $k_2^3$, so $J_2 = 0$ if $\ps(k)/k^3 \rightarrow 0$
as $k \rightarrow 0$.
This is the sharper condition on how strongly large-scale power is
suppressed which was anticipated in Section~\ref{sec:harmonic}. It requires
that $\ps(k)$ cuts off on long lengthscales faster than $k^3$.
If this does not occur, then the integral diverges. (There is a marginal
case where $\ps(k)/k^3$ tends to a finite limit as $k$ approaches
zero. We assume that this case is not physically relevant.)

The $J_2$ integral contains a $\diracd$-function
$\diracd(\vect{k}_2+\vect{k}_3)$. It can therefore be interpreted as
counting contributions to the bispectrum which come from
a correlation between the modes $\vect{k}_2$ and $\vect{k}_3$, in a background
created by $\vect{k}_1$, which exited the horizon in the asymptotic past.
As we have already argued, modes of this sort are included in the FRW
background around which we perturbed to obtain
the correlation functions of $\R$, so we can anticipate that its contribution
should be zero, as the above analysis shows explicitly. In this interpretation,
the condition $\ps(k)/k^3 \rightarrow 0$ as $k \rightarrow 0$ is the condition
that the perturbation does not destroy the FRW background.
Indeed, fluctuations on very large scales in effect describe
transitions from one FRW world to another via a shift in the zero-momentum
modes of the background metric. In this case, there is only one such mode,
which is the scale factor $a(t)$.
These transitions are rather like changing
the vacuum state in a quantum field theory. As a result, fluctuations of
a large volume of the universe between one FRW state and another
are strongly suppressed.

For the case of fluctuations on the scale of a Hubble volume, therefore,
the probability distribution should be written
\begin{equation}
  \label{hubble:prob}
  \Prob(\fluct) = \frac{1}{\sqrt{2\pi} \Sigma} \left[ 1 -
  \left( \frac{\fluct^3}{\Sigma^6} - 3 \frac{\fluct}{\Sigma^4} \right) J
  \right]
  \exp\left( - \frac{\fluct^2}{2\Sigma^2} \right) ,
\end{equation}
where we have used the fact that the corrections are odd in $\fluct$,
and therefore do not contribute to the overall normalization of
$\Prob(\fluct)$. The (dimensionless) coefficient $J$ is
\begin{equation}
  \label{hubble:j}
  J = \int \frac{\d^3 k_1 \, \d^3 k_2 \, \d^3 k_3}{96\pi^2 \prod_i k_i^3
  \window^{-1}(k_i)}
  \diracd(\triangle) \bar{\ps}^2 \A .
\end{equation}
This expression is remarkably simple. Indeed, although the explicit
expression \eref{hubble:j} is preferable for calculation, it can be recast
directly as the integrated bispectrum with respect to $\window$:
\begin{equation}
  \fl
  J = \frac{1}{48(2\pi)^3(2\pi^2)^3} \int \d^3 k_1 \, \d^3 k_2 \, \d^3 k_3 \;
  \langle \R(\vect{k}_1) \R(\vect{k}_2) \R(\vect{k}_3) \rangle
  \window(k_1) \window(k_2) \window(k_3)
\end{equation}
  
As a consistency check, we note that the expectation of
$\fluct$, defined by $\Expect(\fluct) = \int \fluct
\Prob(\fluct) \, \d \fluct$ satisfies $\Expect(\fluct) = 0$. This
is certainly necessary,
since the universe must contain as many underdense regions
as overdense ones, but is a non-trivial restriction, since both the
$\fluct$ and $\fluct^3$ corrections to $\Prob(\fluct)$ do not separately
average to zero. The particular combination of coefficients in
\eref{hubble:prob} is the unique correction [up to $\Or(\fluct^3)$,
containing only odd powers of $\fluct$] which
maintains $\Expect(\fluct)=0$.

Finally, we note that Eqs.~\eref{hubble:prob}--\eref{hubble:j} do not
explicitly involve the cut-off $\Lambda$, except as a limit of integration
and in quantities such as $\Sigma^2_\Lambda$
and $\window_\Lambda$ which possess
a well-defined, finite limit at large $\Lambda$.
As a result, there is no obstruction to taking the $\Lambda \rightarrow \infty$
limit to remove the regulator entirely.

\subsection{When is perturbation theory valid?}
It is known from explicit calculation that the bispectrum is of order
$\ps^2$ multiplied by a small quantity, $\fnl$, which is predicted to be small
when slow-roll is valid. It is therefore reasonable to suppose that whenever
the window functions $\window$ are peaked around some probe wavenumber
$k_{\star}$, one has the approximate relations (on order of magnitude)
$\Sigma^2 \sim \ps_{\star}$ and $J \sim \ps^2_{\star}$, where $\ps_{\star}$
represents the spectrum evaluated at $k = k_{\star}$. Since the $\fluct^3$
correction dominates for $\fluct > \sqrt{3} \Sigma$, this means that for
$\fluct$ not too large, $\fluct \ll \ps_{\star}^{-3/2}$, the perturbative
correction we have calculated will be small. As $\fluct$ increases,
so that $\fluct \gg \ps_{\star}^{-3/2}$, perturbation theory breaks down
and the power series in $\fluct$ needs resummation. In any case,
at such high values of
$\fluct$, the calculation described above ought to be
supplemented by new physics which can be expected to become important at high
energy density. The details of these corrections presumably do not matter too
much, because at any finite order,
the fast-decaying exponential piece suppresses any contributions
from large values of $\fluct$.

At some value of $\fluct$, corrections coming from the
trispectrum can be expected to become comparable to those coming from
the bispectrum that we have computed. Let us assume that for intermediate values
of $\fluct$, such that $\Sigma \ll \fluct \ll \ps_{\star}^{-3/2}$, a contribution
of the form $\fluct^4 \Sigma^{-8} \Jtri$ dominates the correction from
the trispectrum, where $\Jtri$ is the integrated trispectrum with respect to
$\window$.
The trispectrum contribution will be subdominant provided
\begin{equation}
  \frac{\fluct^3}{\Sigma^6} J \gtrsim \frac{\fluct^4}{\Sigma^8} \Jtri ,
\end{equation}
which is true whenever $\fluct < \Sigma^2 J/\Jtri$. The trispectrum from
slow-roll inflation has not yet been explicitly computed, but it is expected
to be proportional to three powers of $\ps$. Therefore, $\Sigma^2 J \sim
\ps_{\star}^3$ and $\Jtri \sim \ps_{\star}^3$ can be expected to be of
roughly equal orders of magnitude, up to a numerical coefficient which can
estimated to be $J/\Jtri \simeq 16\pi \sim 50$. (This number comes from
$4!/3! = 4$ and a factor $4\pi$ from the extra angular integrations which arise
when integrating the trispectrum.) Thus for values of $\fluct$
of order unity, it is reasonable to expect the correction from the bispectrum
to dominate the correction from the trispectrum and higher correlation
functions. However, we caution that if the trispectrum is anomalously large,
or for much larger values of $\fluct$, higher-order terms in the perturbation
series will become relevant.

\section{The probability density function for $\spect(k)$}
\label{sec:probrho}
The probability density function for $\spect(k)$ can be obtained by
a reasonably straightforward modification of the above argument,
taking account of the fact that the constraint, Eq.~\eref{strength:local},
is now a functional constraint. This means that when splitting the functional
measure $[\d \Rsm]$ into a product of $[\d \spect(k)]$ and the orthogonal
degrees of freedom $[\d \Rsm^\perp]$, the result after integrating out the
$\Rsm^\perp$ coordinates gives a functional probability density
in $[\d \spect(k)]$.
In particular, the $\diracd$-function
is now represented as
\begin{equation}
  \int [\d z] \; \exp \left[ \imag \int \d k \; z(k) \left(
  \sum_{n=1}^\infty a^0_{0|n} k^3 \psi_n(k) - \frac{(2\pi)^3}{\sqrt{4\pi}}
  \spect(k) \right) \right] .
\end{equation}
In order to carry out this calculation, we write $z(k)$ formally as
\begin{equation}
  z(k) = \sum_{n=1}^\infty \frac{k^2}{\ps(k) \window^2(k)}
  z_n \psi_n(k) .
\end{equation}
The integration measure $\int [\d z]$ becomes $\prod_n \breve{\mu}
\int_{-\infty}^ \infty \d z_n$, where, as before, $\breve{\mu}$
is a field-independent Jacobian representing the change of variables from
$z(k) \mapsto z_n$. Its value is not relevant to the present calculation.
In addition, we introduce a set of coefficients $\tilde{\spect}_n$ to describe
$\spect(k)$,
\begin{equation}
  \label{nonlocal:rhocoeff}
  \frac{\spect(k)}{k^3} = \sum_{n=1}^\infty \tilde{\spect}_n \psi_n(k) .
\end{equation}
The $\tilde{\spect}_n$ can be calculated using the rule
$\tilde{\spect}_n = \int_{0}^\Lambda \d k \; k^2 \ps^{-1}(k) \window^{-2}(k)
\spect(k) \psi_n(k)$. Note that in order to do so, we have made the implicit
assumption that $\spect(k)/k^3 \rightarrow 0$ as $k\rightarrow 0$, in order
that \eref{nonlocal:rhocoeff} is compatible with the boundary conditions
of the $\psi_n(k)$. We again see the suppression of power in modes with low
$k$.

With these choices, the constraint $\diracd$-function
becomes
\begin{equation}
  \prod_n \breve{\mu} \int_{-\infty}^\infty \d z_n \;
  \exp \left[ \imag \sum_{m=1}^\infty \left( a^0_{0|n} z_n -
  \frac{(2\pi)^3}{\sqrt{4\pi}} z_n \tilde{\spect}_n \right) \right] ,
\end{equation}
As opposed to the nonlocal case of $\fluct$, where a single extra
integration over $z$ coupled to $\fluct$, we now have a situation where
a countably
infinite tower of integrations over $z_n$ couple to the the coefficients
$\tilde{\spect}_n$. In all other respects, however, this calculation is now
much the same as the nonlocal one, and can be carried out in the same way.
The shift of variables necessary to decouple $a^0_{0|n}$ and $z_n$ is
\begin{equation}
  a^0_{0|n} \mapsto a^0_{0|n} + \imag 2\pi^2 (2\pi)^3 z_n ;
\end{equation}
and the shift necessary to decouple the $z_n$ and $\tilde{\spect}_n$ is
\begin{equation}
  z_n \mapsto z_n - \frac{\imag \tilde{\spect}_n}{2\pi^2 \sqrt{4\pi}} .
\end{equation}

When only the two-point function is included, we obtain a Gaussian in the
$\tilde{\spect}_n$,
\begin{equation}
  \label{gauss:spectral}
  \Prob[\spect(k)] \propto \exp\left( - \frac{1}{2} \sum_n \tilde{\spect}_n^2
  \right) .
\end{equation}
The sum over the $\tilde{\spect}_n$ can be carried out using the
completeness and orthogonality relations for the $\psi_n(k)$, and
Eq.~\eref{nonlocal:rhocoeff},
\begin{equation}
  \sum_n \tilde{\spect}_n^2 = \int \d \ln k \; \frac{\spect^2(k)}
  {\ps(k)\window^2(k)} .
\end{equation}
Using this expression, and integrating over all $\spect(k)$ which
give rise to a fluctuation of strength $\fluct$, one recovers
the Gaussian probability profile Eq.~\eref{gauss:gauss} with
variance given by Eq.~\eref{total:variance}. This serves
as a consistency check of \eref{gauss:spectral} and \eref{gauss:gauss}.

When the non-Gaussian correction $\correction^{(0)}$ is included,
one again generates a probability density of the form
\begin{equation}
  \Prob[\spect(k)] \propto (1 - K_0 - K_2) \exp \left( - \frac{1}{2}
  \sum_n \tilde{\spect}_n^2 \right) ,
\end{equation}
where $K_2$ is of the same form as $J_2$, and therefore vanishes for the
same reasons; and $K_0$ has the form
\begin{equation}
  \fl
  K_0 = \int \frac{\d^3 k_1 \, \d^3 k_2 \, \d^3 k_3}{96\pi^2
  \prod_i \ps(k_i) \window(k_i)}
  \diracd(\triangle) \bar{\ps}^2
  \left( 3 \frac{\spect(k_1)}{k_1^3} \frac{\ps(k_2)\window^2(k_2)}
  {k_2^5} \diracd(\vect{k}_2 + \vect{k}_3) - \prod_i \frac{\spect(k_i)}{k_i^3}
  \right) .
\end{equation}
The first term contains a $\diracd$-function which squeezes $k_1$ into the
asymptotic past.
It formally vanishes in virtue of our assumption about the behaviour of
$\spect(k)$ near $k=0$, which is implicit in Eq.~\eref{nonlocal:rhocoeff}.
As a result, the total probability density for the fluctuation spectrum
can be written
\begin{equation}
  \label{fluct:prob}
  \Prob[\spect(k)] \propto (1 - K) \exp \left( - \frac{1}{2}
  \int \d \ln k \; \frac{\spect(k)^2}{\ps(k)\window^2(k)} \right) ,
\end{equation}
where $K$ is given by
\begin{equation}
  \label{fluct:k}
  K = - \int \frac{\d^3 k_1 \, \d^3 k_2 \, \d^3 k_3}{96\pi^2
  \prod_i \ps(k_i) \window(k_i)}
  \diracd(\triangle) \bar{\ps}^2 \prod_i \frac{\spect(k_i)}{k_i^3} .
\end{equation}
As before, one can show that this expression is
consistent with Eqs.~\eref{hubble:prob}--\eref{hubble:j}
by integrating over all $\spect(k)$ which
reproduce a total fluctuation of size $\fluct$, after dropping
another term which is squeezed into the asymptotic past owing to the
presence of a $\diracd$-function. This is a non-trivial consistency
check of \eref{fluct:prob}--\eref{fluct:k}.

As in the local case, Eqs.~\eref{fluct:prob}--\eref{fluct:k} are entirely
independent of $\Lambda$ (except as a limit of integration), so the
regulator can be freely removed by setting $\Lambda = \infty$.

\section{Conclusions}
\label{sec:conclude}
In this paper we have obtained the connexion between
the $n$-point correlation functions of the primordial curvature perturbation,
evaluated at some time $t$,
such as $\langle \R(\vect{k}_1) \cdots \R(\vect{k}_n) \rangle$, and
the probability distribution of fluctuations in the spatial configuration
of $\R$. We have obtained an explicit expression for
the probability of a fluctuation of ``size''
$\fluct$ when $\R$ is smoothed
over regions of order the horizon size. This is a probability density in the
conventional sense. In addition, we have obtained
an expression for the probability that $\fluct$ has a spectrum of the form
$\spect(k)$, that is, $\int \d \ln k \, \spect(k) = \epsilon$. (The
mapping $\spect(k) \mapsto \epsilon$ is many-to-one.) This is a functional
probability density, and can potentially be used to identify features in
the fluctuation spectrum near some specific scale, say of wavenumber
$k \simeq k_{\star}$. Our result is independent of statistical reasoning
based on the central limit theorem and provides a direct route to incorporate
non-Gaussian information from the vertices of the effective quantum field
theory of the inflaton into theories of structure formation.

Both these probabilities are Gaussian in the limit where
$\R$ only possesses a two-point connected correlation function. If
there are higher-order connected correlation functions, then $\R$
exhibits deviations from Gaussian statistics, which we have explicitly
calculated using recent determinations of the inflationary three-point
function during an epoch of slow-roll inflation. Our method can be extended
to incorporate corrections from higher connected $n$-point functions to
any finite order in $n$. We have not computed these higher corrections, since
we anticipate that their contribution is subdominant to the three-point
correction (which is already small), and in any case the relevant
4- and higher $n$-point correlation functions are not yet known explicitly.

Our argument is based on a formal decomposition of the spatial configuration
of the curvature perturbation in $\vect{k}$-space into spherical harmonics,
together with
harmonics along the radial $k$ direction. However, we have emphasized that
our results do not depend on the details of this construction, but require
only a minimal set of assumptions or conditions.
These assumptions are: (A) that the power spectrum $\ps(k)$ goes
to zero sufficiently fast on large scales, specifically so that
$\ps(k)/k^3 \rightarrow 0$ as $k \rightarrow 0$.
(In addition, in the case of the fluctuation spectrum, we also require
$\spect(k)/k^3 \rightarrow 0$ as $k \rightarrow 0$.) Such a
condition is certainly consistent with our understanding of large scale
structure in the universe, and within the perturbative approach we are
using, we have
argued that in fact it describes a self-consistency condition which prevents
perturbative fluctuations from destroying the background FRW spacetime.
In addition, we require a second condition (B) that the spatial configuration
$\Rsp$ can be smoothed $\Rsm$ via a window function $\window$ to obtain
a configuration for which $\Rsm \rightarrow 0$ as $k \rightarrow \infty$,
for which it is fair to compare $\Rsm$ to the primordial power spectrum.

In addition to these fundamental assumptions, which relate to the behaviour
of real physical quantities, a large part of the calculation relied on
an auxiliary technical construction. This construction is based on an
artifical compactification of momentum space, implemented by a hard cutoff
$\Lambda$. There is an associated boundary condition on $\Rsm$ at
$k = \Lambda$ which discretizes the harmonics (partial waves)
in $k$. However, in both the non-local (total fluctuation, $\fluct$) and
local (fluctuation spectrum, $\spect(k)$) cases, the final probability
density is independent of both the details of the partial wave construction
and $\Lambda$ (except as a limit of integration).
It is also independent of the choice of the family of
window functions $\window_\Lambda(k;\kmax)$, and depends only on the limit
$\lim_{\Lambda \rightarrow \infty} \window_\Lambda(k;\kmax) = \window(k;\kmax)$.
Therefore the regulator can be removed by sending
$\Lambda \rightarrow \infty$.
Moreover, the boundary condition at $k = \Lambda$
becomes irrelevant in this limit, which is a familiar result from the theory
of Sturm--Liouville operators. As a consistency check, one can integrate
$\Prob[\spect(k)]$ with the condition $\int \d \ln k \, \spect(k) = \fluct$
in order to obtain $\Prob(\fluct)$.

We conclude by relating our result to earlier work by Ivanov
\cite{ivanov} and Bullock \& Primack \cite{bullock-primack}, who studied
the effect of non-Gaussianities from inflation on PBH formation, but obtained
(na\"{\i}vely) opposite results. In Ref. \cite{ivanov}, it was found that
non-Gaussianities reduce the probability for PBH formation, whereas in
Ref. \cite{bullock-primack}, using a different formalism, the formation
probability was found to be enhanced.
Consider the probability density for the total fluctuation, $\fluct$.
The non-Gaussian correction conserves probability, so it can either
act by moving
probability from the tails into the central region,
or by moving probability from the central region into the tails.
(This effect was also observed in Refs. \cite{bernardeau-uzan,
bernardeau-uzan-a}.)
In the former case, the probability of large fluctuations---and hence
the fraction of the universe going into large collapsed objects---is diminished;
in the latter case, this situation is reversed.
Exactly which occurs depends on the sign of the
coefficient $J$. When $J>0$, the volume of the central region is increased
at the expense of the tails. When $J<0$, the volume of probability in the
tails is increased at the expense of the central region. The crossover
from central region to tails occurs at a threshold amplitude
$\epsilon_{\star} = \sqrt{3} \Sigma$. A priori, it would appear that $J$
can have either sign, depending on the model of inflation under
consideration. This goes some way towards reconciling the apparently
divergent results of Refs. \cite{bullock-primack,ivanov}.

\ack
DS is supported by PPARC.
JCH gratefully acknowledges financial support from the Mexican Council 
for Science and Technology (CONACYT) Studentship No. 179026.
We thank Bernard Carr and Jim Lidsey for useful conversations.

\section*{\refname}

\providecommand{\href}[2]{#2}\begingroup\raggedright\endgroup

\end{document}